\documentclass[11pt]{article}
\usepackage{titlesec}
\titleformat{\paragraph}[runin]{\normalfont\itshape}{\theparagraph.}{.3em}{}[.]\titlespacing{\paragraph}{0pt}{1ex plus .1ex minus .2ex}{.5em}
\usepackage{amsthm,amsmath,amssymb,bm}
\usepackage{mathabx}
\usepackage[T1]{fontenc}
\usepackage[utf8]{inputenc}
\usepackage{lmodern}
\usepackage{mathtools}
\usepackage{dsfont}
\pdfoutput=1
\usepackage[english]{babel}
\usepackage[letterpaper, hmargin=1in, top=1in, bottom=1.2in, footskip=0.6in]{geometry}
\usepackage{graphicx} 
\usepackage{booktabs} 
\usepackage{color}
\definecolor{aquamarine}{rgb}{0.5, 1.0, 0.83}
\definecolor{ao(english)}{rgb}{0.0, 0.5, 0.0}
\definecolor{armygreen}{rgb}{0.29, 0.33, 0.13}
\definecolor{awesome}{rgb}{1.0, 0.13, 0.32}
\definecolor{ballblue}{rgb}{0.13, 0.67, 0.8}
\definecolor{bittersweet}{rgb}{1.0, 0.44, 0.37}
\definecolor{blue}{rgb}{0.0, 0.0, 1.0}
\definecolor{brinkpink}{rgb}{0.98, 0.38, 0.5}
\definecolor{ballblue}{rgb}{0.13, 0.67, 0.8}
\definecolor{brightturquoise}{rgb}{0.03, 0.91, 0.87}
\definecolor{blue-green}{rgb}{0.0, 0.87, 0.87}
\definecolor{caribbeangreen}{rgb}{0.0, 0.8, 0.6}
\definecolor{cyan}{rgb}{0.0, 1.0, 1.0}
\definecolor{amber(sae/ece)}{rgb}{1.0, 0.49, 0.0}
\graphicspath{ {images/} }
\definecolor{vdarkred}{rgb}{0.6,0,0.2}

\definecolor{vdarkred}{rgb}{0.6,0,0.2}
\definecolor{vdarkblue}{rgb}{0,0.2,0.6}
\usepackage[pdftex, colorlinks, linkcolor=vdarkblue,citecolor=vdarkred]{hyperref}

\usepackage[utf8]{inputenc}
\usepackage{lmodern}

\DeclareMathOperator{\Op}{Op}

\newcommand{\id}{\mathbf{1}}

\newcommand{\nn}{\mathbb{N}}
\newcommand{\rr}{\mathbb{R}}
\newcommand{\cc}{\mathbb{C}}

\renewcommand{\d}{{\rm d}}
\newcommand{\e}{{\rm e}}
\renewcommand{\i}{{\rm i}}

\newtheorem{theorem}{Theorem}[section]
\newtheorem{definition}[theorem]{Definition}
\newtheorem{proposition}[theorem]{Proposition}

\author{Tristan Benoist, Martin Fraas, J\"urg Fr\"ohlich}

\title{The appearance of particle tracks in detectors - II\\
the semi-classical realm}

\begin{document}

\maketitle

\vspace{1em}

\begin{abstract}
The appearance of tracks, close to classical orbits, left by charged quantum particles propagating 
inside a detector, such as a cavity periodically illuminated by light pulses, is studied for a family of idealized models. 
In the semi-classical regime, which is reached when one considers highly energetic particles, we present a
detailed, mathematically rigorous analysis of this phenomenon. 
If the Hamiltonian of the particles is quadratic in position- and momentum operators, as in the examples 
of a freely moving particle or a particle in a homogeneous external magnetic field, we show how symmetries, 
such as spherical symmetry, of the initial state of a particle are broken by tracks consisting of infinitely 
many approximately measured particle positions and how, in the classical limit, the initial position and 
velocity of a classical particle trajectory can be reconstructed from the observed particle track.
\end{abstract}

\section{Description of the problem, heuristic considerations, survey of results}\label{Intro}

The purpose of this paper is to provide a partial answer to a fundamental question: \textit{How and under 
what conditions does a classical image of the world emerge from a quantum-mechanical description of reality?} 

The specific phenomenon we propose to analyze is the appearance of \textit{particle tracks} in a cavity periodically illuminated by 
laser pulses or in a cloud chamber, elaborating on results described in \cite{BBFF}. 
We will focus our attention on the example of a highly energetic, charged quantum particle, such as an $\alpha$-particle 
or an electron, whose approximate position is measured periodically by illuminating the region of physical space wherein
it propagates with a pulse of light of wave length $\approx\lambda$. We will assume $\frac{hc}{\lambda}$ to be small as 
compared to the kinetic energy of the particle. The light scattered off the particle is supposed to hit an array of 
photomultipliers. Devices hit by photons then fire with a certain positive probability, an event resulting in a projective state 
reduction in the quantum-mechanical state space of the photomultipliers, which can be recorded. The process described 
here serves to repeatedly determine the approximate position of the charged particle with a precision of $\mathcal{O}(\lambda)$, 
at times $t_n= n\tau,\, n=0,1,2, \dots,$ where $\tau$ is the time elapsing between two subsequent illuminations. 
In between two such indirect approximate measurements of the position of the charged particle
its state is assumed to evolve according to a \textit{Schr\"odinger equation}. 
We propose to show that the approximate positions of a highly energetic particle measured 
at times $t_n$, as described above, and its approximate velocities inferred therefrom lie close to points on a trajectory 
in phase space that is a solution of some \textit{classical Hamiltonian equations of motion}, i.e., the particle positions 
\textit{``track''} a classical orbit. We will consider particles propagating in suitably regular external potentials. But
our main results are formulated for freely moving particles and particles in a homogeneous external 
magnetic field and/or under the influence of a harmonic potential. 

While in an earlier paper \cite{BBFF} we have studied the appearance of particle tracks in an idealized model 
that can essentially be solved exactly, in the present paper we consider fairly general models. In order to be able 
to derive reasonably explicit, mathematically rigorous results, we will study these models in a 
\textit{semi-classical regime}, which is reached when the expected de Broglie wave length of the 
particle is much smaller than the wave length $\lambda$ of the light pulses used to track the position 
of the particle. 

Interesting results on the emergence of particle tracks in detectors, as well as historical remarks 
on various treatments of this phenomenon, can be found in two papers \cite{FT1, FT2}, which 
have proven to be very useful for the work reported in \cite{BBFF} and in the present paper. 
We also draw the readers' attention to paper \cite{Steinmann} where particle tracks have been studied 
within axiomatic quantum field theory.

\subsection{Quantum mechanics of a charged particle, semi-classical regimes}
The Hilbert space of pure state vectors of a charged particle with non-relativistic kinematics and 
without spin (to simplify matters) is given by 
\begin{equation}\label{Hilbert space}
\mathcal{H}_P := L^{2}(\mathbb{R}^{d}, d^{d}x),
\end{equation}
where $\mathbb{R}^{d}\equiv \mathbb{R}^{d}_{x}$ is the configuration space of the particle, and $d^{d}x$ 
is the Lebesgue measure on $\mathbb{R}^{d}$. We will set $d=3$ throughout this introduction.
In between two illuminations by light pulses, the dynamics of the particle is generated by a Hamilton operator, 
$H$, acting on $\mathcal{H}_P$ given by
\begin{equation}\label{Hamiltonian}
H := \frac{1}{2M}\big[P- eA(X)\big]^{2} + gV(X),
\end{equation}
where $M$ is the mass and $e$ the electric charge of the particle (to be kept fixed in what follows), 
$A$ is the vector potential of a \mbox{c-number} external magnetic field (for simplicity chosen to be 
time-independent), $gV$ is an external potential, with $g$ a coupling constant, $P$ is the momentum 
operator, and $X$ is the position operator of the particle. In the Schr\"odinger representation
$$P\Psi(x) = - i \hbar \nabla_{x} \Psi(x), \qquad X\Psi(x) = x\Psi(x), \quad x\in \mathbb{R}^{3}_{x},$$ 
where $\Psi$ is the wave function of the particle. The canonical commutation relations between position- 
and momentum operators are given by
\begin{equation}\label{CCR}
\big[X_i, P_j \big] = i \hbar \delta_{ij} \mathbf{1}, \qquad \big[X_i, X_j\big]=\big[P_i, P_j\big]=0,
\end{equation}
where $\mathbf{1}$ is the identity operator on $\mathcal{H}_P$.
The Schr\"odinger equation for the time dependence of the wave function, $\Psi$, of the particle is given by
\begin{equation}\label{Schrodinger}
i\hbar \frac{\partial}{\partial t} \Psi_{t}= H \Psi_{t}.
\end{equation}
As argued by \textit{Schr\"odinger} \cite{Schrod} and, with more mathematical precision, by 
\textit{Hepp} \cite{Hepp}, the classical limit of this quantum-mechanical description of a charged particle is reached when 
$\hbar \rightarrow 0$, for wave functions that are superpositions of coherent states of the form 
$$\Psi(x)= \exp\Big[\frac{i}{\hbar} (\mathfrak{x}\cdot P-\mathfrak{p}\cdot X)\Big] \exp[-\frac{1}{2} (x/\Lambda)^{2}], \quad (\mathfrak{x}, \mathfrak{p})\in \mathbb{R}^{6},$$
with $\hbar/\Lambda$ kept constant.

In the \textit{Heisenberg picture}, one can establish a \textit{Egorov-type theorem} that says that

\hspace{1.2cm} \textit{ Time Evolution and Quantization commute, up to error terms of $\mathcal{O}(\hbar)$.}

For a precise statement of Egorov's theorem we refer the reader to \cite{BR} where a presentation in the usual setting of semi-classical analysis is given. In Appendix A, we provide a simple proof in a different setting; see Proposition A.5.
In Nature, the value of Planck's constant, $\hbar$, is fixed, and we will henceforth use units in which $\hbar=1$.
Before starting to discuss the main topic of this paper, we propose to identify semi-classical regimes in parameter space,
which are equivalent, mathematically, to a regime corresponding to a very small value of $\hbar$.
Two such regimes are of interest in the context of this paper and will be featured in our analysis.
\begin{enumerate}
\item{We consider a particle with a very large mass $M:=\varepsilon^{-1}m$, with $0<\varepsilon \ll 1$ and 
$m= \mathcal{O}(1)$. We introduce a re-scaled momentum operator (proportional to the \textit{velocity} operator),
$p':= \varepsilon P$, and we set $x':=X$. Then
\begin{equation}\label{CCR'}
\big[x'_i, p'_j\big] = i \varepsilon\, \delta_{ij}\mathbf{1}, \quad \text{ other commutators vanishing,}
\end{equation}
and the Schr\"odinger equation reads
$$i\frac{\partial}{\partial t} \Psi_t = \Big[\varepsilon(2m)^{-1} \big(\varepsilon^{-1}p'- eA(x')\big)^{2}+ g V(x')\Big] \Psi_t.$$
Choosing the vector field $A$ to be large, namely $A= \varepsilon^{-1} A_0$, and the coupling constant to be given by 
$g= \varepsilon^{-1} g_0$, with $A_0$ and $g_0$ kept fixed -- which, physically, means that the \textit{acceleration} 
of the heavy particle is of $\mathcal{O}(1)$, as $\varepsilon$ tends to 0 -- and multiplying both sides of the 
equation by $\varepsilon$, we find that
$$i\varepsilon \frac{\partial}{\partial t}\Psi_t = \Big[\frac{1}{2m} \big(p'- eA_0(x')\big)^{2} + g_0 V(x')\Big] \Psi_t.$$
Comparing this equation and equation \eqref{CCR'} to \eqref{Hamiltonian}, \eqref{Schrodinger} and \eqref{CCR},
we see that $\varepsilon$ plays the role of $\hbar$, and the semi-classical regime apparently corresponds to choosing 
very small values of $\varepsilon$, or, in other words, considering a very heavy particle and preparing it in a state with the property that its typical speed is $\mathcal{O}(1)$.}
\item{Alternatively, we may consider a particle with a mass $M:=m$ of $\mathcal{O}(1)$ prepared in an initial 
state $\Psi_0$ with the property that 
\begin{equation}\label{high-speed}
\langle \Psi_0, \big(P-eA\big)^{2} \Psi_0 \rangle = \mathcal{O}(\varepsilon^{-1}),\quad \text{ with } \,\, 0< \varepsilon \ll 1,
\end{equation}
i.e., the average \textit{kinetic energy} of the particle in its initial state is $\mathcal{O}(\varepsilon^{-1})$, with 
$\varepsilon \ll 1$. We re-scale momentum and position operators as follows:
\begin{equation}\label{rescale}
P=: \varepsilon^{-1/2} p'', \qquad X=:\varepsilon^{-1/2} x''.
\end{equation}
We then have that
\begin{equation}\label{CCR''}
\big[x''_i, p''_j \big] = i\varepsilon\, \delta_{ij} \mathbf{1}, \quad \text{ other commutators vanishing.}
\end{equation}
We choose the vector potential $A\equiv A_{\varepsilon}$ and the potential $V\equiv V_{\varepsilon}$ to depend 
on the variable $\varepsilon$ in such a way that
\begin{equation}\label{A and V}
eA_{\varepsilon}(\varepsilon^{-1/2}x'') \sim \varepsilon ^{-1/2}eA_{0}(x''),\quad \text{ and } \quad 
gV_{\varepsilon}(\varepsilon^{-1/2} x'') \sim \varepsilon ^{-1} g_{0}V_{0}(x''), 
\end{equation}
as $\varepsilon \searrow 0$. In three dimensions ($d=3$), the relation between $A\equiv A_{\varepsilon}$ and $A_0$ 
is automatically fulfilled for a vector potential describing a uniform magnetic field, $B \in \mathbb{R}^{3}$, i.e., for 
$A(x)= \frac{1}{2} (x \times B)$; and the relation between $V\equiv V_{\varepsilon}$ 
and $V_{0}$ is automatically fulfilled for a harmonic potential, e.g., $V(x)= \vert x \vert^{2}$, and $g= g_0$. 
If the relations in~\eqref{A and V} hold, the Schr\"odinger equation reads
$$i \varepsilon \frac{\partial}{\partial t} \Psi_t = \Big[\frac{1}{2m}\big(p'' - e A_{0}(x'')\big)^{2} + 
g_{0}V_{0}(x'') \Big] \Psi_t\,.$$
As above, inspecting Eq.~\eqref{CCR''} and this particular form of the Schr\"odinger equation, we find that the 
variable $\varepsilon$ plays the role of Planck's constant $\hbar$. Apparently, the semi-classical regime 
corresponds to taking a very small value of $\varepsilon$, i.e., preparing an initial state with a very large average kinetic energy, $\mathcal{O}(\varepsilon^{-1})$, and then re-scaling the momentum and  position operators accordingly.}
\end{enumerate}

\noindent
{\bf{The semi-classical regime:}} Regions 1 and 2 in parameter space may be treated in a unified way. For this purpose, we set 
\begin{equation}\label{CCR'''}
 \hat{p}= p', \,\,\hat{x}:=x'=X, \quad \text{ or } \quad\hat{p}=p'',\,\, \hat{x}:=x'', \quad \text{ with } 
 \big[\hat{x}_i, \hat{p}_j\big] = i \varepsilon\, \delta_{ij}\mathbf{1},
 \end{equation}
(other commutators vanishing). Dropping the subscript ``0'' on $A, V$ and $g$, we consider the Schr\"odinger equation
\begin{align}\label{Schrod}
i \varepsilon \frac{\partial}{\partial t}\, \Psi_t = H_{P}\, \Psi_t\,,  \quad\text{where }\quad H_{P}:=  \frac{1}{2m}\big(\hat{p} - eA(\hat{x})\big)^{2} + gV(\hat{x}).
\end{align}
The semi-classical regime corresponds to values $\varepsilon \ll 1$ and initial wave functions, $\Psi_0$, with 
the properties that $\Psi_0 \in \mathcal{H}_{P}, \Vert \Psi_0 \Vert =1$, and
\begin{equation}\label{Var}
 \Delta_{\Psi_0}\hat{x} \cdot \Delta_{\Psi_0} \hat{p} = \mathcal{O}(\varepsilon),
 \end{equation}
 where, as usual,
 $$\Delta_{\Psi_0}A:= \sqrt{\langle \Psi_0, \big(A- \langle A \rangle_{\Psi_0}\big)^{2} \Psi_0 \rangle} \quad \text{ with } \quad
\langle A\rangle_{\Psi_0}:= \langle \Psi_0, A \Psi_0 \rangle. $$

We propose to study the dynamics of the quantum particle in the semi-classical regime described by Eqs.~\eqref{CCR'''}, 
\eqref{Schrod} and \eqref{Var} and to analyze the effect of repeated approximate particle-position measurements, 
taking place every $\tau= \mathcal{O}(1)$ seconds, on the propagation of the particle. The particle position, $\hat{x}$, 
is measured approximately by scattering light with a wave length $\approx \lambda$ off the particle; 
($\lambda$ is taken in the same units as $\vert \hat{x}\vert$). We assume that $\lambda$ is much 
larger than the average de Broglie wave length of the particle in the state $\Psi_0$. In between two consecutive approximate 
position measurements the wave function of the particle is assumed to propagate according to the Schr\"odinger 
equation \eqref{Schrod}.

\subsection{Approximate particle-position measurements}\label{particle position}
Next, we sketch a crude model describing approximate measurements of the particle position; see also \cite{BBFF}.  
We imagine that, every $\tau$ seconds, a pulse of light is emitted into the region of 
physical space $\mathbb{R}^{3}$ where the charged particle is located, and the light scattered by the particle is caught 
by an array of photomultipliers that fire with a positive probability when hit by scattered photons. The firing of 
photomultipliers represents an event whose effect is taken into account by applying the state reduction postulate; 
see Eq.~\eqref{state}, below.  Let $\mathfrak{H}$ denote the Hilbert space of pure state vectors of the array of 
photomultipliers. This space contains a distinguished vector, $\Omega_0$, that corresponds to \textit{quiescent} 
photomultipliers.
Furthermore, there is an operator, $Q=(Q_1, \dots, Q_k),$ with commuting components, $Q_i$, $i=1, ..., k,$ acting 
on $\mathfrak{H}$, which has a discrete spectrum, $\sigma(Q)$, given by, for instance, a (subset of
a) $k$-dimensional lattice. The vector $\Omega_0$ is the eigenvector of $Q$ corresponding to an eigenvalue 
denoted by $q_{\infty} \in \sigma(Q)$. Points $q=(q_1, ..., q_k) \in \sigma(Q)$ correspond to certain subsets of 
photomultipliers. The firing of the photomultipliers indexed by a point $q \in \sigma(Q)$ is correlated with the 
event that the position of the charged particle is somewhere within a distance of $\mathcal{O}(\lambda)$ of 
a point $\mathfrak{x}(q) \in \mathbb{R}^{3}$ uniquely determined by $q$; ($q_{\infty}$ indicating that the
 charged particle has escaped to a region not illuminated by the light pulses). 
Let $\mathfrak{H}_q \subset \mathfrak{H}$ denote the eigenspace of the operator $Q$ corresponding 
to the $k$-tuple $q$ of eigenvalues of $Q$. The event corresponding to the firing of the photomultipliers
indexed by a point $q\in \sigma(Q)$ is represented by the orthogonal projection operator, $\pi_{q}$, onto the eigenspace 
$\mathfrak{H}_q$ of $Q$ corresponding to the eigenvalues $q$. One has that
$$\sum_{q\,\in\, \sigma(Q)} \pi_{q} = \mathbf{1}_{\mathfrak{H}}.$$
In every eigenspace $\mathfrak{H}_{q}$ we may choose an orthonormal basis of eigenvectors, $\varphi_{q, \alpha}$, 
labelled by the eigenvalue $q$ and an additional index $\alpha=1,2, \dots$. We denote by 
$\pi_{q,\alpha}= \vert \varphi_{q, \alpha}\rangle \langle \varphi_{q, \alpha}\vert $ the orthogonal projection onto 
$\varphi_{q, \alpha}$, and we have that $\sum_{\alpha} \pi_{q, \alpha} = \pi_q$.

We assume that, after firing and the recording of an event $\pi_{q}$, $q\in \sigma(Q)$, the photomultipliers relax 
back to the quiescent state $\Omega_0$, with a relaxation time, $T$, much shorter than the time, $\tau$, elapsing between 
two consecutive light pulses. Furthermore, we assume that the time elapsing between the release of a light pulse, 
the subsequent firing of the photomultipliers and the recording of an event $\pi_q$ is sufficiently short 
that the motion of the charged particle can be neglected during this process. 

Let $\rho$ be a \textit{density matrix} on $\mathcal{H}_P$, i.e., a non-negative trace-class operator on $\mathcal{H}_P$ with 
$\text{tr}(\rho) = 1$, encoding the state of the particle just \textit{before} a firing of the photomultipliers caused by light scattering 
and the recording of an event $\pi_{q}$. We are interested in determining the \textit{Born probability} 
of the event $\pi_{q}$. Let $U$ be the unitary operator describing the evolution of the initial state, 
\mbox{$\rho \otimes | \Omega_0\rangle \langle \Omega_0 |$}, of the \textit{total} system (consisting of the charged particles, 
the radiation field and the photomultipliers) at the instant when a light pulse is emitted to the state reached \textit{after} 
light scattering by the charged particle, but just \textit{before} the event $\pi_{q}$ is recorded. 
Since the state $\rho$ of the charged particle is assumed to be very nearly \textit{constant} during this process, 
the operator $U$ has the form
\begin{equation}\label{Q-meas}
\big[U\big(\rho \otimes | \Omega_0\rangle \langle \Omega_0 |\big)U^{*}\big](x,y) = \rho(x,y) \cdot U(x) 
| \Omega_0\rangle \langle \Omega_0| U(y)^{*}, \quad x, y \text{ in } \,\sigma(\hat{x})=\mathbb{R}^{3},
\end{equation}
where $\rho(x,y)$ is the operator kernel of $\rho$ in the Schr\"odinger representation, $\hat{x}$ is the position operator of 
the particle, $\sigma(\hat{x})$ denotes the spectrum of $\hat{x}$, and the operators $U(x)$  are unitary operators on 
$\mathfrak{H}$, for all $x\in \sigma(\hat{x})$. 
The $x$-dependence of the operator $U(x)$ entangles the state of the charged particle with the state of the 
photomultipliers; (the radiation field does not have to be taken into account explicitly). The Born probability we are looking 
for is given by a functional, $\Pi(q\vert \cdot)$, on the space of density matrices given by
\begin{equation}\label{Born}
\Pi(q \vert \rho) := \text{tr}_{\mathcal{H}_{P}\otimes \mathfrak{H}} \Big[U(\hat{x})(\rho \otimes |\Omega_0\rangle \langle \Omega_0 |)U(\hat{x})^{*} \pi_{q}\Big]\,.
\end{equation}
We can write this functional as 
\begin{equation}\label{amplitudes}
\Pi(q \vert \rho) = \sum_{\alpha} \text{tr}_{\mathcal{H}_P} \Big[\widehat{f}_{q, \alpha}(\hat{x})\, \rho\, \widehat{f}_{q,\alpha}(\hat{x})^{*}\Big]\,,
\end{equation}
where the \textit{(transition) amplitudes} $\widehat{f}_{q, \alpha} \equiv \widehat{f}_{q,\alpha}(\hat{x})$ are multiplication 
operators corresponding to multiplication by the functions
\begin{equation}\label{amplitude}
{f}_{q,\alpha}(x):= \langle \varphi_{q, \alpha}, U(x) \Omega_{0}\rangle, \quad \text{for }\,x\in \sigma(\hat{x}),
\end{equation}
with $\varphi_{q, \alpha}$ the eigenvector in the range of the projection $\pi_{q,\alpha}$. Thus
$$
{f}_{q, \alpha}(x)^{*}\cdot {f}_{q, \alpha}(x) = \langle U(x) \Omega_{0}, \pi_{q, \alpha} U(x) \Omega_{0}\rangle.
$$
We note that 
\begin{equation}\label{sumrule}
\sum_{q, \alpha} {f}_{q, \alpha}(x)^{*}\cdot {f}_{q, \alpha}(x) = \Vert U(x)\Omega_{0}\Vert^{2}= 1,\,\quad i.e., \quad \,
\sum_{q, \alpha} \widehat{f}_{q, \alpha}^{*} \cdot \widehat{f}_{q, \alpha} = \mathbf{1}_{\mathcal{H}_P}\,.
\end{equation}
Since the point $q\in \sigma(Q)$ is supposed to track the position $x\in \mathbb{R}^{3}$ of the charged particle, 
we assume that the amplitudes $f_{q, \alpha}(x), x\in \sigma(\hat{x}) = \mathbb{R}^{3},$ are of rapid decrease 
in the quantity $\vert \mathfrak{x}(q)-x \vert/\lambda$, where the map 
$\mathfrak{x}: \sigma(Q)\ni q \mapsto \mathfrak{x}(q) \in \mathbb{R}^{3}$, introduced above, 
maps a $k$-tuple $q$ of eigenvalues of $Q$ to a unique approximate particle position $\mathfrak{x}(q)$.

Let $\rho$ be the density matrix on $\mathcal{H}_{P}$ encoding the state of the charged particle 
right \textit{before} a firing of the photomultipliers and $\tau$ seconds before the next light pulse is emitted. 
We propose to determine the state, $\rho(q)$, of the particle \textit{after} the firing of the photomultipliers, assumed to
correspond to the point $q\in \sigma(Q)$, and just \textit{before} the next light pulse is emitted. Assuming that the Born
probability $\Pi(q \vert \rho)$ does not vanish, this state is given by
\begin{align}\label{state}
\rho \mapsto \rho(q):= \frac{\Phi_{q}^{*}\big(\rho\big)}{\text{tr}_{\mathcal{H}_P}
\big[\Phi_{q}^{*}\big(\rho\big)\big] }, \qquad
 \Phi_{q}^{*}\big(\rho\big)\equiv \Phi_{\varepsilon, q}^{*}(\rho) := 
e^{-i\tau H_P/\varepsilon}\,\sum_{\alpha}\Big(\widehat{f}_{q, \alpha}\, \rho\, \widehat{f}_{q, \alpha}^{*}\Big)\, 
e^{i\tau H_P/\varepsilon}\,.
\end{align}
Identity \eqref{sumrule} implies that the map $\rho \mapsto \sum_{q\in \sigma(Q)} \Phi^{*}_{q}(\rho)$ is completely 
positive and trace-preserving. The map $\rho \mapsto \rho(q)$ can be iterated to determine a state, 
$\rho(q_0, q_1, \dots , q_n)$, of the particle after $n+1$ firings of photomultipliers:
\begin{align}\label{track record}
\rho(q_0, q_1, \dots, q_n)= \frac{\Phi_{q_n}^{*} \circ \cdots \circ \Phi_{q_0}^{*}(\rho_0)}{\text{tr}_{\mathcal{H}_P}\big[\Phi_{q_n}^{*} \circ \cdots \circ \Phi_{q_0}^{*}(\rho_0)\big]}\,,
\end{align}
where $\rho_0 $ is the initial state of the particle.
With a measurement record $\big\{q_0, q_1, \dots, q_n\big\}$ we associate the quantity
\begin{equation}\label{proba}
\mathbb{P}_{\varepsilon, \rho_0}^{(n)}(q_0, q_1, \dots, q_n) =\text{tr}_{\mathcal{H}_P}\big[\Phi_{\varepsilon, q_n}^{*} \circ \cdots \circ \Phi_{\varepsilon, q_0}^{*}(\rho_0)\big],
\end{equation}
which is non-negative. By Eq.~\eqref{sumrule} we have that 
$$\sum_{q\in \sigma(Q)} \text{tr}_{\mathcal{H}_P}\big[\Phi_{q}^{*}(\rho)\big] = \text{tr}_{\mathcal{H}_P}\big[\rho\big] =1.$$
It thus follows that
\begin{align}\label{Kolmogorov}
\sum_{q_n\in \sigma(Q)} \mathbb{P}_{\varepsilon, \rho_0}^{(n)}(q_0,q_1, \dots, q_n) &= \mathbb{P}_{\varepsilon, \rho_0}^{(n)}(q_0,q_1, \dots, q_{n-1}), \quad \text{hence}\nonumber\\
\sum_{q_j\in \sigma(Q),\, j=0,1,\dots, n} \mathbb{P}_{\varepsilon, \rho_0}^{(n)}(q_0,&q_1, \dots, q_n) =
\text{tr}_{\mathcal{H}_P}[\rho_0]=1\,,
\end{align}
for an arbitrary density matrix $\rho_0$ on $\mathcal{H}_P$. Thus, 
$\mathbb{P}_{\varepsilon, \rho_0}^{(n)}(q_0,q_1, \dots, q_n)$ can be interpreted as the \textit{probability of the 
measurement record} $\underline{q}_n := \big\{q_0, q_1, \dots, q_n \big\}$, conditioned on the initial state of the particle being
given by $\rho_0$. 

From this point on, the quantum mechanics of the radiation field and of the photomultipliers does not play a significant 
role, anymore. It is subsumed completely in Eqs.~\eqref{state} and \eqref{track record}. In conventional jargon, the 
so-called \textit{``Heisenberg cut''} may apparently be placed at the level of the approximate particle-position 
measurements described by the rule \eqref{state}, which conforms to standard lore.

The goal of our paper is to show that, for $\varepsilon \ll 1$, i.e., in the semi-classical regime, 
the measurement record corresponds to a sequence of approximate particle positions, 
$\big\{\mathfrak{x}(q_j)\,\big|\, j=0,1,...,n\big\}$, lined up near a particle orbit corresponding to a solution of
classical equations of motion determined by a Hamilton function, $h_P$, whose quantization is the Hamilton 
operator $H_P$. Moreover, we show that, for simple particle dynamics, the random initial data of the particle, and 
hence the entire particle orbit, can be reconstructed from a large number of measurements of approximate particle 
positions, i.e. when $n \gg 1$. This follows ideas outlined in the introduction of \cite{BBFF}. 

Let 
\begin{equation}\label{phase space}
\Gamma:= \mathbb{R}^{3}_x \oplus \mathbb{R}^{3}_p, \qquad h_P(x, p):= \frac{1}{2m}\big(p-eA(x)\big)^{2} + gV(x)
\end{equation}
denote the classical phase space of the charged particle and the Hamilton function corresponding to the Hamilton
operator $H_P$ introduced in \eqref{Schrod}, respectively. The Hamilton function $h_p$ generates a
symplectic flow, $(\phi_t)_{t\in \mathbb{R}}$, on $\Gamma$, where $t$ denotes time, and we write
$(x_t, p_t) := \phi_{t}(x,p)$. Given a function $a(x,p)$ on $\Gamma$ belonging to the Schwartz space, 
$\mathcal{S}(\Gamma)$, let $\widehat{a}(\hat{x}, \hat{p})$ denote the operator obtained from $a(x,p)$ by 
\textit{Weyl quantization}; (see Sect.~2). We will invoke a Egorov-type theorem that says that
\begin{equation}\label{Egor}
e^{itH_{p}/\varepsilon} \,\widehat{a} \, e^{-itH_{P}/\varepsilon} = \widehat{a \circ \phi_t} + \mathcal{O}(\varepsilon).
\end{equation}
Furthermore, if $a$ and $b$ belong to $\mathcal{S}(\Gamma)$ then
\begin{equation}\label{commutator}
\Vert \big[\widehat{a}, \widehat{b}\big] \Vert= \mathcal{O}(\varepsilon).
\end{equation}
These facts will be discussed in Sect.~2 and proven for a certain class of functions $a$ in Appendix A. Applying Eqs.~\eqref{Egor} and \eqref{commutator} to
the right side of \eqref{proba}, we find that
\begin{equation}\label{proba'}
\mathbb{P}_{\varepsilon, \rho_0}^{(n)}(q_0, q_1, \dots, q_n) = \text{tr}\Big[\rho_0\, \prod_{j=0}^{n} \Big(\sum_{\alpha} 
\widehat{f}_{q_j, \alpha}(\widehat{x}_{j\tau})^{*} \cdot \widehat{f}_{q_j, \alpha}(\widehat{x}_{j\tau})\Big) \Big]+ 
\mathcal{O}(\varepsilon)\,.
\end{equation}
Choosing a family $\big\{ \rho_{0, \varepsilon}\big\}_{\varepsilon >0}$ of density matrices indexed by $\varepsilon$ 
with the property that their \textit{Wigner quasi-probability distributions} converge to a probability measure, 
$\text{d}\mu_{0}$, on phase space $\Gamma$, it follows that
\begin{equation}\label{asymptotics}
\mathbb{P}_{\varepsilon, \rho_{0, \varepsilon}}^{(n)}(q_0, q_1, \dots, q_n) = \int_{\Gamma} \prod_{j=0}^{n} \Big(\sum_{\alpha} 
\big| f_{q_j, \alpha}(x_{j\tau})\big|^{2}\Big) 
\text{d}\mu_{0}(x,p) + \mathcal{O}(\varepsilon)\,, \quad\text{ as } \,\, \varepsilon \rightarrow 0\,,
\end{equation}
where $x_{j\tau}$ is the configuration space coordinate of the phase-space point 
$\phi_{j\tau}(x,p)$. The error term $\mathcal{O}(\varepsilon)$ in \eqref{asymptotics} is expected to grow rapidly 
in the number, $n+1$, of approximate particle position measurements. However, in deriving our results we only
require convergence, as $\varepsilon \rightarrow 0$, for arbitrary finite values of $n$.

Assuming that $A(x) = \frac{1}{2}(x \times B)$, where $B$ is a uniform external magnetic field independent of time, and that
the potential $V(x)$ is harmonic, we can use standard arguments from statistics to show that the expression on the right 
side of Eq.~\eqref{asymptotics} is peaked on measurement records 
$\big\{q_0,q_1, \dots, q_n\big\}$ corresponding to classical particle orbits $\{x_t \,|\,t= j\tau, j=0, 1, \dots, n\}$. More precisely,
if the amplitudes $f_{q, \alpha}(x)$ are of rapid decrease in $\vert \mathfrak{x}(q)-x \vert/\lambda$ the points
$\mathfrak{x}(q_j)\in \mathbb{R}^{3}$ are typically within a distance of $\mathcal{O}(\lambda)$ of the points $x_{j\tau}$, 
for all $j=0,1, \dots, n$, where $(x_t, p_t) = \phi_{t}(x,p)$ solves the classical Hamiltonian equations of motion 
with initial conditions $(x_0,p_0)$. Moreover, the probability that the initial conditions $(x_0,p_0)$ belong to a cell 
$\Delta$ of phase space $\Gamma$ is given by $\mu_{0}(\Delta)$ (i.e., \textit{Born's Rule} holds in the limiting regime
where $\varepsilon \ll 1$).

Precise statements of our main results are presented in Sect.~3. Detailed proofs are contained in Sects.~4 
and 5. In Sect.~6, some concrete examples of particle dynamics are sketched along the lines of the discussion 
in \cite{BBFF}. Preliminaries, concerning, e.g., Weyl quantization etc., are discussed in Sect.~2. 
Some technical proofs are presented in Appendix A.\\

\textit{Acknowledgements:} We are grateful to \textit{Detlev Buchholz} for some comments on our paper \cite{BBFF} and 
for drawing our attention to a paper by the late \textit{Othmar Steinmann}, where the problem of particle 
tracks in detectors has been studied within the formalism of axiomatic quantum field theory, see \cite{Steinmann}. We thank
\textit{Miguel Ballesteros} and \textit{Baptiste Schubnel} for earlier collaboration on somewhat related problems and \textit{Philippe Blanchard} for numerous discussions on quantum mechanics.
T.B. would like to thank \textit{Fabrice Gamboa} for his advice on the estimation results and \textit{Jean-Marc Bouclet} for pointing out appropriate references for the semi-classical analysis results we use. The research of T.B. has been supported by ANR project ESQUISSE (ANR-20-CE47-0014-01) of the French National Research Agency (ANR). The research of T.B. and M.F. has been supported
by ANR project QTraj (ANR-20-CE40-0024-01) of the French National Research Agency (ANR).

\section{Weyl quantization, repeated indirect measurements} \label{quantization} 
In order to imbed the ideas presented between Eqs.~\eqref{phase space} and \eqref{asymptotics} of Sect.~1 
into precise mathematics, we need to recapitulate some basics concerning Hamiltonian dynamics and the 
process of quantization. In the following we will make use of \textit{Wigner-Weyl quantization} of classical Hamiltonian 
systems with finitely many degrees of freedom. We follow conventions inspired by \cite[\S 8.4]{DerGerard}. 
The phase space, $\Gamma$, is taken to be the one introduced in \eqref{phase space}, i.e.,
\begin{equation}\label{ps}
\Gamma:= \mathbb{R}^{d}_x \oplus \mathbb{R}^{d}_p.
\end{equation}
Points of $\Gamma$ are denoted by Greek letters $\xi, \zeta, \dots$ Let $\mathcal{S}(\Gamma)$ be the Schwartz 
space of test functions on $\Gamma$. The Fourier transform, $\mathcal{F}(a)$, of a function 
$a\in \mathcal{S}(\Gamma)$ is defined by
\begin{equation}\label{FT}
\mathcal{F}(a)(\zeta)\equiv \tilde{a}(\zeta):= (2 \pi)^{-2d}\int_{\Gamma} a(\xi) \,e^{i \xi^{t}\Omega\zeta} \text{d}\xi\,,
\end{equation}
where the superscript $t$ indicates transposition, and $\Omega$ is the $2d\times 2d$ matrix given by
\begin{equation}\label{Omega}
\Omega = \begin{pmatrix} 0& -\mathbf{1}_{d}\\ \mathbf{1}_{d}& 0 \end{pmatrix}.
\end{equation}
For a positive number $\varepsilon \in (0, \varepsilon_0]$, the \textit{Weyl quantization}, $\text{Op}_{\varepsilon}(a),$ 
of an arbitrary function $a\in \mathcal{S}(\Gamma)$ is defined, formally, by
\begin{equation}\label{WQ}
\text{Op}_{\varepsilon}(a)\equiv \widehat{a}:= \int_{\Gamma} \tilde{a}(\zeta)\, W(\zeta)\, \text{d}\zeta\,,
\end{equation}
where
\begin{equation}\label{W Op}
W(\zeta)\equiv W_{\varepsilon}(\zeta) :=\text{exp}\big[i(\zeta^{t}\Omega\widehat{\xi}\,)\big], \qquad \widehat{\xi}:= \begin{pmatrix} \hat{x} \\ \hat{p} \end{pmatrix},
\end{equation}
are the usual \textit{Weyl operators}, and $\hat{x}$ and $\hat{p}$ are the position and the momentum operator, respectively, on the Hilbert space $\mathcal{H}_P$, which satisfy the Heisenberg commutation relation; see Eq.~\eqref{CCR'''}, Sect.~1.
The Weyl operators $W(\zeta), \zeta\in \Gamma,$ are unitary and satisfy the 
\textit{Weyl (commutation) relations}
\begin{equation}\label{WR}
W(\zeta_1)\,W(\zeta_2) = e^{-i\frac{\varepsilon}{2}\,\zeta_{1}^{t} \Omega\,\zeta_2} W(\zeta_1 +\zeta_2).
\end{equation}
We observe that 
\begin{equation}\label{Wstar}
W(\zeta)^{*} = W(-\zeta) \quad \text{and }\,\, W(0)=\mathbf{1}.
\end{equation}
\\
\noindent
{\bf{Digression on the mathematical meaning of Eq.~\eqref{WQ}:}} In order to render the definition of the operation of quantization, $\text{Op}_{\varepsilon}$, 
more precise, we introduce a sesquilinear form, 
$B_{\varepsilon}(a |\cdot, \cdot)$, on $\mathcal{H}_P \times \mathcal{H}_P$ given by
\begin{equation}\label{sesqui}
B_{\varepsilon}(a | \Phi, \Psi):= \int_{\Gamma} \tilde{a}(\zeta) \,\langle \Phi, W(\zeta)\,\Psi \rangle\, \text{d}\zeta\,,
\quad \Phi, \Psi \text{ in } \mathcal{H}_P\,.
\end{equation}
Since $W(\zeta)$ is unitary, for arbitrary $\zeta \in \Gamma$, we have that 
$\vert \langle \Phi, W(\zeta) \Psi \rangle\vert  \leq \Vert \Phi\Vert\cdot \Vert \Psi \Vert$; hence 
\begin{equation}\label{eq:norm_ineq_B(a)}
\vert B_{\varepsilon}(a | \Phi, \Psi) \vert \leq \Vert \tilde{a} \Vert_{1} \, \Vert \Phi \Vert \cdot \Vert \Psi \Vert
\end{equation}
where $\|\cdot\|_1$ denotes the $L^1$-norm.
Furthermore, the function $\zeta \mapsto \langle \Phi, W(\zeta)\Psi \rangle$ is continuous in $\zeta$, for arbitrary $\Phi$ and 
$\Psi$ in $\mathcal{H}_P$, and, since the selfadjoint operator $\mathfrak{x}\cdot \hat{p}-\mathfrak{p}\cdot \hat{x}$
has purely absolutely continuous spectrum, for arbitrary $0\neq \zeta=(\mathfrak{x}, \mathfrak{p})^{t} \in \Gamma$, it tends 
to $0$, as $\vert \zeta \vert \rightarrow \infty$, by the Riemann-Lebesgue lemma. The Riesz representation theorem thus implies that there is a
 unique bounded operator, $\text{Op}_{\varepsilon}(a)$, on $\mathcal{H}_P$ with the property that
\begin{equation}\label{Op}
B_{\varepsilon}(a | \Phi, \Psi)= \langle \Phi, \text{Op}_{\varepsilon}(a)\Psi\rangle, \quad \forall \, \Phi, \Psi \,\text{ in } \,\mathcal{H}_P.
\end{equation}
By Equation~\eqref{eq:norm_ineq_B(a)}, $\|\Op_\epsilon(a)\|\leq \|\tilde a\|_1$.
Thanks to this norm-bound and the continuity properties of the function 
$\zeta \mapsto \langle \Phi, W(\zeta)\Psi \rangle$, the operation $\text{Op}_{\varepsilon}$ of quantization 
can be extended to a larger function space, in the following denoted by $S$, strictly containing 
$\mathcal{S}(\Gamma)$. Examples of $S$ are the space of functions that are inverse Fourier transforms 
of finite complex Borel measures on $\Gamma$, or the space of bounded $C^{\infty}$-functions on $\Gamma$ with bounded derivatives. 
We equip $S$ with a norm, $\Vert (\cdot) \Vert$, with the property that 
$\Vert \text{Op}_\varepsilon(a)\Vert\leq \Vert a \Vert, \,\forall \, a \in S.$ In the following, the function space $S$ is
assumed to conform to the following definition.\\

\noindent {\bf{Definition 2.1.}}$\mathbf{(S)}$ \textit{The space $S$ is an $\varepsilon$-independent, normed function 
space contained in the space of bounded measurable functions on $\Gamma$. It contains the set of Schwartz functions $\mathcal S(\Gamma)$. When equipped with point-wise multiplication and complex conjugation $S$ is a normed $^{*}$-algebra with a sub-multiplicative norm.}

\textit{For an arbitrary function $a\in S$, the sesquilinear from $B_{\varepsilon}(a|\cdot, \cdot)$ introduced in \eqref{sesqui} is 
well defined on $\mathcal{H}_P\times \mathcal{H}_P$, with
$$\underset{0<\varepsilon\leq \varepsilon_0}{\text{sup}}\, \Vert B_{\varepsilon}(a |\Phi, \Psi) \Vert \leq \Vert a \Vert\,
 \Vert \Phi\Vert\cdot \Vert \Psi \Vert, \quad \text{for arbitrary }\,\,\, \Phi, \Psi \,\text{ in }\, \mathcal{H}_P.$$ 
The operator $\text{Op}_{\varepsilon}(a)$ is the unique bounded operator satisfying \eqref{Op}. Its operator norm is dominated by the $S$-norm of $a$:
$\Vert Op_{\varepsilon}(a) \Vert \leq \Vert a \Vert, \forall a \in S$.}
\textit{We finally require that,
for arbitrary functions $a$ and $b$ in $S$,}
\begin{equation}\label{class lim}
\underset{\varepsilon \searrow 0}{\text{lim}}\,\, \Vert \text{Op}_{\varepsilon}(a)\, \text{Op}_{\varepsilon}(b) - 
\text{Op}_{\varepsilon} (a\cdot b) \Vert = 0.
\end{equation}
 
 \noindent{\bf{Remark 2.2.}} \textit{The choices}
 $$S:=\big\{ a\in C^{\infty}(\Gamma)\, \big|\, \text{sup}_{\xi \in \Gamma} \vert \partial^{\alpha} a(\xi)\vert < \infty, \, \forall \, \text{multi-indices }\, \alpha\big\},$$
\,\, \textit{and}
 $$S:= \big\{a \,\big| \, a = \mathcal{F}^{-1} (\tilde{a}), \, \text{where }\,\tilde{a}(\text{d}\zeta) \text{ is a finite complex Borel measure on }\, \Gamma \big\}$$
 \vspace{0.15cm}
\textit{ alluded to above, are convenient for our purposes; see \cite{Zworski}, § 4, and  Appendix A.} \hfill{\large{$\square$}}

Norms on $S$ are specified in \cite{Zworski} (Theorem~4.23), and in Appendix A, respectively.
 In the following we will often use the short-hand notation
$$\widehat{a}:= \text{Op}_{\varepsilon} (a), \quad a\in S.$$
By \eqref{WQ} and \eqref{Wstar}, we have that
$$\widehat{a}^{*} =\int_{\Gamma} \overline{\tilde{a}(\zeta}) W(-\zeta)\,\text{d}\zeta.$$
 If the function $a$ is real then $\overline{\tilde{a}(\zeta)}=\tilde{a}(-\zeta)$, and we  conclude that 
 $$\widehat{a}^{*} = \widehat{a} \,\,\text{ is self-adjoint}, \, \text{for an arbitrary \textit{real} element }\,\, a\in S.$$
 
 Next, we return to considering the \textit{dynamics of the particle}. We suppose that the classical Hamilton function $h_P$ 
 is as specified in Eq.~\eqref{phase space} of Sect.~1, for a smooth potential $V$ on $\Gamma$, with $\partial^{\alpha}V$ 
 bounded, for $\vert \alpha \vert \geq 2$; (one may allow $V$ to also depend on $p$). This function does not belong to the 
 space $S$; but the appropriate  quantization, $H_P\equiv \widehat{h}_P$, of $h_P$ has already been introduced 
 in Eq.~\eqref{Schrod} of Sect.~1 (see also \cite{DerGerard}, \S 8.4). Under natural conditions on $V$ (see, e.g., \cite{BS}), the Hamiltonian $H_P$ 
 is a self-adjoint operator on $\mathcal{H}_P$. 
 Hence 
 \begin{equation}\label{propagator}
 U_{\varepsilon} := \text{exp}\big[-i \tau H_P/\varepsilon \big]
 \end{equation}
 is a unitary operator, for arbitrary $\varepsilon \in (0, \varepsilon_0]$;  (the parameter $\tau$ is the time elapsing 
 between two consecutive approximate particle-position measurements, as discussed in Sect.~1). The classical 
 Hamilton function $h_P$ determines a symplectic flow $(\phi_t)_{t\in \mathbb{R}}$ on $\Gamma$. 
 We define a symplectomorphism $\phi: \Gamma \rightarrow \Gamma$ by setting $\phi:= \phi_{\tau}$. 
  
We require $U_{\varepsilon}$ and $\phi$ to satisfy the following semiclassical approximation assumption.

\vspace{0.15cm}\noindent {\bf{Assumption (SC).}} \textit{We assume that the symplectomorphism $\phi$ preserves the space $S$, 
in the sense that $a\circ \phi \in S, \,\forall \, a\in S$, and, moreover, that}
\begin{equation}\label{Egorov}
\text{lim}_{\varepsilon \searrow 0}\, \Vert U_{\varepsilon}^{*} \text{Op}_{\varepsilon}(a)U_{\varepsilon} - 
\text{Op}_{\varepsilon}(a\circ \phi) \Vert =0\,,
\end{equation}
\textit{ for all functions} $a\in S$.\\

\noindent{\bf{Remark 2.3.}} \textit{If $S$ is chosen to be the space of smooth bounded functions on $\Gamma$, as in 
Remark 2.2, and for a Hamiltonian $H_P$ as specified above, Assumption (SC) is a consequence of Egorov's theorem
(see, e.g., \cite{BR}, Theorem 1.2). A short proof of Assumption (SC) is provided in Appendix A for a suitably chosen 
function space $S$ under a somewhat abstract condition on the flow $(\phi_t)_t$. Here we just remark that if the Hamiltonian is quadratic in $\hat{x}$ and $\hat{p}$, i.e., 
if the particle dynamics is quasi-free, then it follows that}
\begin{equation}\label{symp matrix}
U_{\varepsilon}^{*} \text{Op}_{\varepsilon}(a) U_{\varepsilon} = \text{Op}_{\varepsilon}(a\circ \phi_J), \quad \text{ for arbitrary }\,\varepsilon>0,
\end{equation}
\textit{where $\phi_{J}(x,p) = J\begin{pmatrix} x \\ p \end{pmatrix},$ with $J$ a symplectic matrix on $\mathbb{R}^{2d}$, i.e.,
$J^{t} \Omega J = \Omega,$ where $\Omega$ is the matrix introduced in \eqref{Omega}.
This choice of Hamiltonian covers the examples of a freely moving particle, a particle in a constant external 
magnetic field and the harmonic oscillator. (It is discussed in detail in \cite{BBFF} and in Sect.~6).}

\subsection{Indirect approximate measurements of particle position}\label{measurement}
Next, we put the analysis of approximate particle-position measurements presented in Subsection 
\ref{particle position} into a slightly more abstract guise. Let $E$ be a locally compact metric space equipped with its Borel $\sigma$-algebra, and let d$\nu$ be a 
$\sigma$-finite measure on $E$. A weak measurement of some particle properties, using a suitable 
instrument, can be described by an operator (belonging to the algebra $\text{Op}_{\varepsilon}(S)$) corresponding to what, in 
Subsect.~\ref{particle position}, has been called an \textit{amplitude}, $f_{\alpha}$, which is defined as follows: Let
$f_{\alpha}: E\times \Gamma \rightarrow \mathbb{C}$, $\alpha = 1,2, \dots$, be measurable functions with the following properties: 

\vspace{0.15cm}\noindent{\bf{Properties of amplitudes.}}
\begin{enumerate}
\item[(P1)]{For d$\nu$-almost all points $q\in E$, $f_{q, \alpha}: \xi \mapsto f_{q, \alpha}(\xi)$ is an element of the space $S$, and, 
for an arbitrary continuous compactly supported function $\psi$ on $E$,
$$\langle \psi \rangle_{\alpha}: \xi \mapsto \langle \psi\rangle_{\alpha}(\xi) :=\int_{E} \psi(q) \vert f_{q, \alpha}(\xi) \vert^{2} \text{d}\nu(q) \,\text{ belongs to the space }\, S, \quad \forall \alpha.$$}
\item[(P2)]{The functions $E\ni q \mapsto \Vert f_{q,\alpha} \Vert^{2}_\infty$ are locally integrable with respect to d$\nu$ 
and summable in $\alpha$. (For simplicity, we will henceforth assume that the number of indices $\alpha$ is bounded
 by some finite integer, $N_0$, for all $q\in E$.)}
\item[(P3)]{Let $\widehat{f}_{q,\alpha} \equiv \text{Op}_{\varepsilon}(f_{q, \alpha}), q\in E,$ be the quantization of the functions 
$f_{q, \alpha}$ (which, by (P1), is well defined for almost every $q\in E$). Then
\begin{equation}\label{tracepres}
\int_{E} \sum_{\alpha}\Big(\widehat{f}_{q, \alpha}^{*}\,\widehat{f}_{q,\alpha}\Big) \,\text{d}\nu(q) = \mathbf{1}_{\mathcal{H}_P},
\end{equation}
for arbitrary $\varepsilon \in (0, \varepsilon_0]$, or, equivalently,}
$$\int_E\sum_{\alpha}|f_{q,\alpha}(\xi)|^2\text{d} \nu(q)=1, \text{ for any }\, \xi\in \Gamma. $$
\end{enumerate}
{\bf{Remark 2.4.}} \textit{(1) Property (P3) guarantees that the map}
$\rho \mapsto \int_{E} \sum_{\alpha}\big(\widehat{f}_{q, \alpha} \rho \widehat{f}_{q, \alpha}^{*}\big)\,\text{d}\nu(q)$ 
\textit{is completely positive and trace-preserving. Properties (P1) and (P2) are tailored to the use of semi-classical 
analysis, as described below. We note that if the first half of property (P1) holds then the second half of
(P1) follows for our examples of spaces $S$.}

\textit{(2) If the functions $f_{q, \alpha}, q\in E,$ are \textit{independent} of the momentum variable $p$ then properties
(P2) and (P3) hold, provided the functions $q\mapsto \Vert f_{q, \alpha}\Vert$ are locally square-integrable and, 
for every $\xi\in \Gamma$, $q\mapsto \sum_{\alpha} \vert f_{q, \alpha}(\xi) \vert^{2}$ is a probability density with respect to the measure} $\text{d}\nu$.

\textit{(3) Condition (P3) ensures that for any $\xi\in \Gamma$, }$\sum_{\alpha}|f_{q,\alpha}(\xi)|^2\text{d}\nu(q)$
\textit{ is the law of a random variable $Q$ taking values in $E$.}

Let $\rho$ be a density matrix on $\mathcal{H}_P$ representing the state of the particle right before a weak 
measurement of a particle property, as described by amplitudes $\widehat{f}_{q,\alpha}$, is made, as discussed 
in Subsect.~1.2. The probability distribution of the measurement result $Q=q\in E$ is given 
by the Born probabilities introduced in Eq.~\eqref{Born} of Subsect.~1.2, namely\footnote{Note that, in the following, an abstract random variable is denoted by a capital letter, while its values are denoted by the corresponding lower-case letter. Example: the approximate position of a particle is a random variable denoted by $Q$, its measured values are denoted by $q$.}
$$ Q \sim \Pi(q \vert \rho)\text{d}\nu(q):=\sum_{\alpha}\text{tr}\big(\widehat{f}_{q, \alpha}\,\rho\, \widehat{f}_{q, \alpha}^{*}\big)\text{d}\nu(q)$$
When conditioned on a measurement outcome $Q=q$, with $\Pi(q\vert \rho) >0$, the state, $\rho(q)$, of the particle 
$\tau$ seconds after the measurement, right before the next measurement, is given by
\begin{equation}\label{q-state}
\rho(q):= \frac{\Phi_{\varepsilon, q}^{*}\big(\rho\big)}{\text{tr}_{\mathcal{H}_P}\big[\Phi_{\varepsilon, q}^{*}(\rho\big)\big] },
\end{equation}
where the maps $\rho \mapsto \Phi_{\varepsilon, q}^{*}(\rho), q\in E,$ are as in Eq.~\eqref{state}, Subsect.~1.1.
Thanks to property (P3), Eq.~\eqref{tracepres}, the map $\rho \mapsto \int_{E} \Phi_{\varepsilon, q}^{*}(\rho) \text{d}\nu(q)$ 
is completely positive and trace-preserving.

The maps $\Phi_{\varepsilon, q}^{*}$ can be iterated. We define 
\begin{equation}\label{iterated state}
\Phi_{\varepsilon,\underline{q}_n}^{*} := \Phi_{\varepsilon,q_n}^{*}\circ \cdots \circ \Phi_{\varepsilon,q_0}^{*},
\end{equation}
with $\underline{q}_{n}:=\big\{q_0, q_1, \dots, q_n\big\}$, for any $n=0, 1,2, \dots$ 
We define a probability measure $\text{d}\mathbb{P}_{\varepsilon,\rho_0}^{(n)}(\underline{q}_n)$ by
\begin{equation}\label{prob-measure}
\text{d}\mathbb{P}^{(n)}_{\varepsilon, \rho_0}(\underline{q}_n):= \text{tr}\big[\Phi^{*}_{\varepsilon, \underline{q}_n}(\rho_0)\big]\, \text{d}\nu^{\otimes (n+1)}(\underline{q}_n)\,,
\end{equation}
where $\rho_0$ is the initial state of the particle; see Eq.~\eqref{proba}. Using \eqref{tracepres} 
(see also \eqref{Kolmogorov}) and \textit{Kolmogorov's extension lemma}, we conclude that there exists a probability measure, 
$\text{d}\mathbb{P}_{\varepsilon, \rho_0}$, on the space, $\mathfrak{Q}:=E^{\times\mathbb{N}_0}$, of infinite sequences 
of measurement outcomes, $\underline{q}_{\infty}= \big(q_n\big)_{n\in \mathbb{N}_0}$, with the property that 
\begin{equation}\label{P-meas}
\text{d}\mathbb{P}_{\varepsilon, \rho_0}\big|_{E^{n+1}}= \text{d}\mathbb{P}^{(n)}_{\varepsilon, \rho_0}\,,
\end{equation}
where $E^{n+1}$ consists of all measurable subsets of $\mathfrak{Q}$ that do not depend on $q_j, j\geq n+1$. 

\section{Survey of results}\label{survey}
In this section we present precise statements of our main results. We begin with a theorem 
that says that, in the semi-classical regime, i.e., for $0<\varepsilon \ll 1$, the process of particle measurements 
described in Subsect.~2.1 is close to a process of independent approximate particle measurements whose laws 
follow a classical particle trajectory determined by the Hamiltonian dynamics generated by 
the Hamilton function $h_P$ introduced in Eq.~\eqref{phase space}. 
The random initial condition, $\xi_0 \in \Gamma$, of the particle trajectory is distributed according 
to a probability measure, $\text{d}\mu_0$, on phase space $\Gamma$ that describes the limiting initial 
state of the particle corresponding to a family of states, $\big\{\rho_{0,\varepsilon}\big\}_{\varepsilon\in (0, \varepsilon_0]}$, as $\varepsilon \searrow 0$; see Eq.~\eqref{asymptotics}, 
Subsect.~1.1, and \eqref{class measure}, below.
\begin{theorem} We assume that the function space $S$ be as required in Definition 2.1, and that the
time-$\tau$ symplectomorphism $\phi$ satisfy Assumption (SC); (see Eqs.~\eqref{class lim} and \eqref{Egorov}, 
and Appendix A). Furthermore, we assume that there exist a probability measure, $\text{d}\mu_0$, on phase space $\Gamma$ such that, for any function $a$ on $\Gamma$ belonging to the space 
$S$,
\begin{equation}\label{class measure}
\underset{\varepsilon \searrow 0}{\text{lim}}\,\, \rho_{0, \varepsilon} \big(\text{Op}_{\varepsilon}(a)\big) =
 \int_{\Gamma} a(\xi)\,\text{d}\mu_{0}(\xi)\,.
\end{equation}
Let $\big(\xi_n\big)_{n\in \mathbb{N}_0}$ be the classical process with $\xi_0 \sim \text{d}\mu_0$ and 
$\xi_n= \phi(\xi_{n-1}), \,\forall \,n>0$. For $\xi \in \Gamma$, let $Q(\xi)$ be the random variable whose law is 
given by
$Q(\xi) \sim \sum_{\alpha}\big| f_{q, \alpha}(\xi)\big|^{2} \text{d}\nu(q)$; and let $\big(Q_n\big)_{n\in \mathbb{N}_0}$ 
be the process whose law is given by the probability measure
$\mathrm{d}\mathbb{P}_{\varepsilon, \rho_0}$, see \eqref{P-meas}. Then
$$ \big(Q_n\big)_{n\in \mathbb{N}_0} \overset{\mathcal{L}}{\underset{\varepsilon \searrow 0}{\longrightarrow}} 
\big(Q(\xi_n)\big)_{n \in \mathbb{N}_0}\,,$$
where the different copies, $Q(\xi_{n}), n=0,1,2, \dots,$ of the random variable $Q(\xi)$ are independent.\, $\square$
\end{theorem}

The proof of Theorem 3.1 is given in the next section.

\vspace{0.15cm}\noindent
{\bf{Remark 3.2}} \textit{Theorem 3.1 can be generalized to open systems, (i.e., systems interacting with an environment).
We suppose that, in the Heisenberg picture, the quantum-mechanical evolution of operators of some open system 
is determined by a unital, completely positive map} \mbox{$\Psi_{\varepsilon} : \mathcal{O} \rightarrow \mathcal{O},$}
\textit{where} $\mathcal{O}$ is the $C^{*}$-\textit{algebra  generated by the operators} 
$\big\{\text{Op}_{\varepsilon}(a)\big| a\in S\big\}$.
\textit{The operator norm of} $\Psi_{\varepsilon}$ \textit{is bounded by $1$; see \cite{RD}. (If the system were isolated, 
as above, the map $\Psi_{\varepsilon}$ would be given by $\Psi_{\varepsilon}(X)=U_{\varepsilon}^{*} \,X\, U_{\varepsilon}$, 
for some unitary operator $U_{\varepsilon}$ on $\mathcal{H}_P$.) We replace Assumption (SC), Eq.~\eqref{Egorov}, 
by the assumption}
$$\underset{\varepsilon \searrow 0}{\text{lim}}\,\, \Vert \Psi_{\varepsilon} \big(\text{Op}_{\varepsilon} (a)\big) - 
\text{Op}_{\varepsilon}\big(\Pi \,a\big) \Vert =0,$$
\textit{where} $\Pi: S\rightarrow S$ \textit{is a Markov kernel. Then Theorem 3.1 holds, provided}
 $(\xi_n)_{n \in \mathbb{N}_0}$ \textit{is chosen to be the Markov chain with kernel} $\Pi$, \textit{and the law of} 
 $\xi_0$ \textit{is given by the probability measure} $\text{d}\mu_0$.


\vspace{0.15cm}
In \cite{L-P}, Examples III.1, III.3 and III.4, the following result clarifying the status of Eq.~\eqref{class measure} has been established.

\vspace{0.2cm}\noindent{\bf{Proposition 3.3.}} \textit{Let $\rho_{0, \varepsilon}:= \big| \psi_{\varepsilon}^{(x_0, v_0)}\rangle \langle \psi_{\varepsilon}^{(x_0, v_0)}\big|$, with 
\begin{equation}\label{wave function}
\psi_{\varepsilon}^{(x_0,p_0)}(x):= \varepsilon^{-(d\beta/2)} h\Big(\frac{x-x_0}{\varepsilon^{\beta}}\Big) \,
e^{i(x\cdot p_0/\varepsilon)},
\end{equation}
where $h \in L^{2}(\mathbb{R}^{d}_{x}, d^{d}x)$, with $\Vert h \Vert_{2}=1$, and $(x_0, p_0)^{t} \in \Gamma$.
Then 
$$\psi_{\varepsilon}^{(x_0,p_0)} \in L^{2}(\mathbb{R}^{d}_{x}, d^{d}x), \,\text{ with }\,\,\, \Vert \psi_{\varepsilon}^{(x_0,p_0)} \Vert_{2}=1, $$
and, for an arbitrary $\beta \in [0,1]$, the limit in Eq.~\eqref{class measure} exists, with} $\text{d}\mu_0$ \textit{given by the following formulae:}
\begin{itemize}
\item{ \textit{For} $\beta =0, \text{d}\mu_{0}(x,p) = \delta(p-p_0) \vert h(x-x_0) \vert^{2} d^{d}x d^{d}p.$}
\item{\textit{For} $\beta=1, \text{d}\mu_{0}(x,p) = \delta(x-x_0) \vert \widetilde{h}(p-p_0) \vert^{2} d^{d}x d^{d}p$, \textit{where
$\widetilde{h}$ is the Fourier transform of $h$}.}
\item{\textit{For} $\beta \notin \{0,1\},  \text{d}\mu_{0}(x,p) = \delta(x-x_0)\,\delta(p-p_0) d^{d}x d^{d}p$. \hspace{5.8cm}$\square$}
\end{itemize}
\noindent
{\bf{Remark 3.4.}} \textit{(1) Whenever $\beta \notin\{0,1\}$, the trajectory $(\xi_n)_{n\in \mathbb{N}_0}$ is deterministic, with 
initial condition $\xi_0=(x_0, p_0)^{t}$.}\\
\textit{(2) For $\beta=1$, and if the function $h$ is invariant under space rotations (an s-wave state), the distribution of initial momenta of the particle is rotation-invariant, too. But the sequence of observed approximate 
particle positions determines a definite initial direction of particle motion, which is, however, random. 
The isotropy of the initial particle state is mirrored in the rotation invariance of the \textit{distribution} on the space 
of particle tracks; but conditioning on a sample track breaks this symmetry. This becomes evident when 
estimating the initial momentum of the particle from the data of the particle track, as we will 
discuss in Section~\ref{sec:estim}.}\\
\textit{(3) Proposition 3.3 extends to mixed states given by}
$$\rho_{\varepsilon} = \int_{\Gamma} |\psi_{\varepsilon}^{(x_0,p_0)}\rangle \langle \psi_{\varepsilon}^{(x_0,p_0)}| \,
\text{d}\lambda(x_0, p_0), $$
\textit{where} $\text{d}\lambda$ \textit{is a probability measure on $\Gamma$. By dominated convergence, the limit in 
Eq.~\eqref{class measure} then exists, with the measure} $\text{d}\mu_0$ \textit{given by} (i) $\text{d}\mu_0 = \text{d}\lambda$, \textit{for} $\beta\notin \{0,1\}$, (ii) $\text{d}\mu_{0}(x, p) = \int_{\mathbb{R}_{x}^{d}} \vert h(x-y)\vert^{2} \text{d}\lambda(y,p)$,
\textit{for} $\beta =0$, \textit{and} (iii) $\text{d}\mu_0 (x, p) = \int_{\mathbb{R}_{p}^{d}} \vert \widetilde{h}(p-r)\vert^{2} 
\text{d}\lambda(x,r)$, \textit{for} $\beta=1$.\\

Next, we consider measurements of the approximate \textit{particle-position} at times $t_n= n\tau, n=0,1,2,\dots$ 
The functions $f_{q, \alpha}(\xi), q\in E,$ are then independent of the momentum variable $p$, and we assume,
merely for simplicity, that the index $\alpha$ only takes a single value, so that it can be dropped. Let
$$\mathfrak{x}: E \rightarrow \mathbb{R}^{d}, \,\,\,E\ni q \mapsto \mathfrak{x}(q) \in \mathbb{R}^{3}$$
be the map introduced in Sect.~1.2, where $\mathfrak{x}(q)$ is interpreted to be the approximate particle position 
corresponding to a measurement of $Q$ with outcome $q\in E$. We focus our attention on approximate particle-position 
measurements.

\vspace{0.15cm}\noindent{\bf{(Space-) Translation-Invariant Instruments.}} \textit{We assume that 
the image of $E$ under the map $\mathfrak{x}$ is given by $\mathbb{R}^{d}_{x}$, and
that the push forward of the measure} d$\nu(q)$ \textit{by the map $\mathfrak{x}$ is 
the Lebesgue measure, $d^{d}x$, on $\mathbb{R}^{d}_{x}$; i.e.,}
\begin{equation}\label{transl-invariance}
\int_{E} F(\mathfrak{x}(q)) \text{d}\nu(q) = \int_{\mathbb{R}^{d}} F(x) d^{d}x,
\end{equation}
\textit{for any $L^1$ function $F$ on $\mathbb{R}^{d}$.
Furthermore, we assume that there exist a function 
\mbox{$g: \mathbb{R}^{d}\rightarrow \mathbb{C}$} belonging to the space $S$ such that 
\mbox{$f_{q}(\xi=(x,p)^{t}) := g(\mathfrak{x}(q)-x)$}.}

\vspace{0.15cm}Theorem 3.1 then takes the following form.

\vspace{0.2cm}
\noindent
{\bf{Corollary 3.5.}}\label{cor:translation_inv}
 \textit{Under the hypotheses of Theorem 3.1, and for translation-invariant instruments,}
$$\big(\mathfrak{x}(Q_n)\big)_{n\in \mathbb{N}_0} \overset{\mathcal{L}}{\underset{\varepsilon \searrow 0}{\longrightarrow}}
\big(x_n + \kappa_n \big)_{n\in \mathbb{N}_0}\,,$$
\textit{where the law of} $\big(Q_n\big)_{n\in \mathbb{N}_0}$ \textit{is given by the probability measure} 
$\text{d}\mathbb{P}_{\varepsilon, \rho_0}$ \textit{defined in \eqref{prob-measure}}, 
$\big((x_n, p_n)^{t}\big)_{n\in \mathbb{N}_0}$ \textit{is the classical process} 
$(\xi_n=\phi_{n\tau}(\xi_0))_{n\in \mathbb{N}_0}$ \textit{introduced in Theorem 3.1, and} 
$(\kappa_n)_{n \in \mathbb{N}_0}$ \textit{is a sequence of independent, 
identically distributed random variables with values in $\mathbb{R}^{d}$ whose law is given by the probability measure} 
$\big| g(\kappa) \big|^{2} d^{d}\kappa$.\\

\smallskip\noindent
{\bf{Remark}} \textit{We can reformulate this corollary as follows. For small values of the deformation 
parameter $\varepsilon$, the sequence of observed approximate particle positions has a distribution 
similar to the one of a perturbation of the classical particle orbit:
$\big\{\mathfrak{x}(Q_0), \mathfrak{x}(Q_1), \dots\big\}\sim\big\{x_0 + \kappa_0, x_1 + \kappa_1, \dots\big\}$,
where $\kappa_0, \kappa_1, \dots $ are independent identically distributed random variables and the points 
$\big\{x_n=x(\tau n)\,\vert\, n=0,1, \dots\big\}$ lie on a classical 
particle orbit $\big(x(t)\big)_{ t\in \mathbb{R}}$ corresponding to some random initial condition $\xi_0\in \Gamma$ 
whose distribution, $\text{d}\mu_0$, is consistent with Born's rule.}

\bigskip
Next, we propose to investigate how particle trajectories $(\xi_n)_{n\in \mathbb{N}_0}$ can be reconstructed 
from sequences of data of approximate particle-position measurements, $(Q_n)_{n\in \mathbb{N}_0}$, in the 
limit where $\varepsilon$ tends to 0. We require the assumptions specified in Corollary 3.5. To simplify our 
notations, we assume that $E= \mathbb{R}^{d}$ and that the map $\mathfrak{x}$ is the identity, i.e., 
$\mathfrak{x}(q)=q, \, \forall q \in \mathbb{R}^{d}$. 

We introduce the measures
\begin{align}\label{mu measure}
\Lambda(\xi, \text{d}q)&:= \big| g(q-x) \big|^{2} d^{d}q, \,\,\,\text{ with }\,\, \xi=\begin{pmatrix} x\\p \end{pmatrix},\\ 
\label{P measure}
&P(\xi_0, \text{d}\underline{q}_{\infty}):= \bigotimes_{n=0}^{\infty} \Lambda(\xi_n, \text{d}q_n), 
\end{align}
where $\xi_n = \phi(\xi_{n-1})$, as in Theorem 3.1. If \eqref{class measure} holds then the measures $\text{d}\mathbb{P}_{\varepsilon, \rho_{0, \varepsilon}}$ 
defined in Eq.~\eqref{P-meas} converge to the measure $P(\mu_{0}, \text{d}\underline{q}_{\infty}):= 
\int_{\Gamma} \text{d}\mu_{0}(\xi_0)\, P(\xi_0, \text{d}\underline{q}_{\infty})$, as $\varepsilon \searrow 0$.

\vspace{0.2cm}\noindent
{\bf{Assumption (QF).}} \textit{(1) The particle dynamics is ``quasi-free'', i.e., the Hamilton function $h_P$ in 
Eq.~\eqref{phase space} of Sect.~1.2 is quadratic in $x$ and $p$. There then exists a symplectic matrix $J$ 
on phase space $\Gamma$ such that $\phi(\xi)=\phi_{J}(\xi)= J\xi,\, \forall\, \xi \in \Gamma.$} \\
\textit{(2) The first and the second moment of the measure} $\Lambda(\xi, \text{d}q)$ \textit{exist, and}
\begin{equation}\label{momen}
\int_{\mathbb{R}^{d}} q\, \Lambda(\xi, \text{d}q)= x, 
\end{equation}
\textit{with $\xi$ and $x$ as in \eqref{mu measure}.}\\

Our aim is to understand how
the particle states, $\xi_n, n=0,1,2, \dots$, and, in particular, the initial condition $\xi_0$ of the particle trajectory, 
can be determined from a given sequence, $\underline{q}_{\infty} =(q_n)_{n\in \mathbb{N}_0}$, of outcomes of approximate
particle-position measurements.

\vspace{0.2cm}\noindent
{\bf{Theorem 3.6.}} \textit{Assume that the hypotheses of Corollary 3.5 are valid and that Assumption (QF) holds. Assume moreover that the classical dynamics has no stable or unstable  manifolds (i.e., $\operatornamewithlimits{spec}J\subset U(1)\equiv \text{exp}(i \mathbb{R})$) and is non trivial in any direction of space (i.e., $(\mathbf{1}_d\,, 0)J\left(\substack{0\\\mathbf{1}_d}\right)$ is invertible).}

\textit{Then, in the classical limit $\varepsilon \searrow 0$, there exists a sequence $(\tilde \xi_n)_{n\in\mathbb N_0}$ 
of measurable functions on $\mathfrak{Q}:= E^{\times \mathbb{N}_0}$ with values in $\Gamma$ such that, for each $n\in \mathbb N_0$, $\tilde\xi_n$ depends only $(Q_0,Q_1,\dotsc,Q_n)$ and with the property that}
$$\lim_{n\to\infty}\mathbb E_{P}(\|\tilde\xi_n-\xi_0\|^2)=0$$
\textit{with $\mathbb E_P$ the expectation with respect to $P(\xi_0,\mathrm{d}\underline{q}_\infty)$ and 
$\Vert \cdot \Vert$ the usual Euclidean norm on $\mathbb{R}^{2d}$. Hence $\tilde \xi_n$ converges to 
$\xi_0$ in probability, as $n$ tends to $\infty$.}

\vspace{0.15cm}More precise statements of this theorem, with an explicit expression for $\tilde \xi_n$, and proofs are 
given in Sect.~\ref{sec:estim}.

\vspace{0.2cm}These somewhat abstract considerations will be illustrated by concrete examples in Sect.~6.

\section{Proofs of Theorem 3.1 and Corollary 3.5}
We begin with the proof of Theorem 3.1, which relies on the fact that 
$\widehat{a}\cdot \widehat{b}= \widehat{a\cdot b} + \mathcal{O}(\varepsilon)$, for arbitrary functions $a$ and $b$ 
belonging to the space $S$; see Eq.~\eqref{class lim}, Definition 2.1. Thanks to property (P1) of amplitudes, stated at the beginning 
of Sect.~2.1, the assumption that the amplitdes $f_{q, \alpha}(\xi)$ belong to the space $S$ (with $\alpha$ taking only 
finitely many values) implies that $\sum_{\alpha} \vert f_{q,\alpha}(\xi)\vert^{2}$ belongs to $S$, too, for $\nu$-almost 
all $q\in E$, because $S$ is assumed to be a $^{*}$-algebra. For every $\xi \in \Gamma$, we define a measure 
$\Lambda$ on the space $E$ by setting
$\Lambda(\xi, \text{d}q):= \sum_{\alpha} \vert f_{q,\alpha}(\xi)\vert^{2} \text{d}\nu(q)$, and we then define the
measure $P(\xi_0, \text{d}\underline{q}_{\infty})$ on $\mathfrak{Q}$ as described 
in Eq.~\eqref{P measure}.
Let $\mathbb{E}_{P}$ denote expectation with respect to the measure $P$, and let 
$\mathbb{E}_{\mathbb{P}_{\varepsilon}}$ denote expectation with respect to the measure 
$\text{d}\mathbb{P}_{\varepsilon}\equiv \text{d}\mathbb{P}_{\varepsilon, \rho_{0, \varepsilon}}$, where 
$\text{d}\mathbb{P}_{\varepsilon, \rho_{0}}$ has been defined in \eqref{P-meas} and the family of states 
$\big\{\rho_{0, \varepsilon}\big\}_{0<\varepsilon \leq \varepsilon_0}$ is chosen such that \eqref{class measure} holds.
We first show that
\begin{equation}\label{convergence}
\mathbb{E}_{\mathbb{P}_{\varepsilon}}(\psi_{0}\big(Q_0) \cdots \psi_{n}(Q_n)\big) 
\underset{\varepsilon \searrow 0}{\rightarrow} \mathbb{E}_{P}\big(\psi_{0}(Q_0) \cdots \psi_{n}(Q_n)\big),
\end{equation}
for arbitrary non-negative, compactly supported continuous functions $\psi_0, \dots, \psi_n$ on $E$. Then, using 
the decomposition of continuous functions into positive and negative parts and the density (in the $L^1$-norm) 
of compactly supported continuous functions in the set of bounded continuous functions, the convergence 
stated in \eqref{convergence} yields Theorem 3.1.

For an arbitrary non-negative, compactly supported continuous function $\psi$ on $E$, the map
$$\Phi_{\varepsilon, \psi}: X \mapsto \int_{E} \psi(q) \Big(\sum_{\alpha} \widehat{f}_{q,\alpha}^{*} U_{\varepsilon}^{*}\, X\, U_{\varepsilon} \widehat{f}_{q, \alpha}\Big) \text{d}\nu(q), \qquad X\in B(\mathcal{H}_P),$$
with $U_{\varepsilon}:= \text{exp}[-i\tau H_{P}/\varepsilon],$ is completely positive, since it is expressed as
a Kraus decomposition. From properties (P1) and (P2) of amplitudes (see Sect.~2.1), 
$$\|\Phi_{\varepsilon,\psi}(\id)\|\leq \int_E\psi(q)\sum_\alpha\|f_{q,\alpha}\|^2\mathrm{d}\nu(q)<\infty.$$
The Russo-Dye Theorem then implies that $\Phi_{\varepsilon,\psi}$ is bounded uniformly in $\varepsilon$;
see Corollary 1 in \cite{RD}. This map is the adjoint of the map 
$\Phi_{\varepsilon, \psi}^{*}(\cdot):=\int_{E}\psi(q) \Phi_{\varepsilon, q}^{*}(\cdot)\text{ d}\nu(q),$ which acts on density matrices,
where $\Phi_{\varepsilon,q}^{*}$ has been introduced in Eq.~\eqref{state}.

Next, we show that, for an arbitrary non-negative, compactly supported continuous function
$\psi$ on $E$ and any $a\in S$,
\begin{equation}\label{limits}
\underset{\varepsilon \searrow 0}{\text{lim}}\, \Vert \Phi_{\varepsilon, \psi}\big(\text{Op}_{\varepsilon}(a)\big) - 
Op_{\varepsilon}(\langle \psi\rangle\, a\circ \phi)\Vert = 0,
\end{equation}
where
\begin{align}
\langle \psi\rangle = &\sum_{\alpha} \langle \psi \rangle_{\alpha}, \,\,\,\text{ with }\nonumber \\
\langle \psi \rangle_{\alpha}(\xi) : =&\int_{E} \psi(q) \vert f_{q, \alpha}(\xi) \vert^{2} \text{d}\nu(q). \label{meanvalue}
\end{align}
By property (P1), Sect.~2.1, we have that $\langle \psi \rangle \in S$. Moreover, Assumption (SC) stated in Sect.~2 
implies that $a \circ \phi \in S$, where $\phi$ is the time-$\tau$ symplectic map on $\Gamma$. 
Then Eq.~\eqref{class lim} in Definition 2.1 entails that $\text{lim}_{\varepsilon \searrow 0} \Vert \text{Op}_{\varepsilon}
(\langle \psi \rangle) 
\text{Op}_{\varepsilon}(a\circ \phi) - \text{Op}_{\varepsilon}(\langle \psi \rangle\,a\circ \phi) \Vert =0. $ 
Thus, \eqref{limits} follows from
$$
\lim_{\varepsilon \searrow 0} \,\Vert \Phi_{\varepsilon, \psi}\big(\text{Op}_{\varepsilon}(a)\big) - \text{Op}_{\varepsilon}
(\langle \psi \rangle) \text{Op}_{\varepsilon}(a \circ \phi)\Vert =0.
$$
Since quantization, i.e., the operation $\text{Op}_{\varepsilon}$, is linear, we have that
\begin{align}
\Phi_{\varepsilon, \psi}\big(\text{Op}_{\varepsilon}(a)\big)& - \text{Op}_{\varepsilon} (\langle \psi \rangle) 
\text{Op}_{\varepsilon}(a \circ \phi) =\nonumber \\
= &\sum_{\alpha}\int_{E} \psi(q) \big[  \widehat{f}^{*}_{q, \alpha} U_{\varepsilon}^{*} 
\text{Op}_{\varepsilon}(a) U_{\varepsilon} \widehat{f}_{q, \alpha} - \text{Op}_{\varepsilon}( \vert f_{q,\alpha}(\xi)\vert^{2}) \text{Op}_{\varepsilon}(a \circ \phi)\big] \text{d}\nu(q)\,.\nonumber
\end{align}
Assumption (SC) and \eqref{class lim} imply that
$$\underset{\varepsilon \searrow 0}{\text{lim}}\, \Vert \widehat{f}^{*}_{q, \alpha} U_{\varepsilon}^{*} 
\text{Op}_{\varepsilon}(a) U_{\varepsilon} \widehat{f}_{q, \alpha} - \text{Op}_{\varepsilon}( \vert f_{q,\alpha}(\xi)\vert^{2}) \text{Op}_{\varepsilon}(a \circ \phi)\Vert =0,$$
for $\nu$-almost every $q$. Furthermore,
$$\Vert \widehat{f}^{*}_{q, \alpha} U_{\varepsilon}^{*} \text{Op}_{\varepsilon}(a) U_{\varepsilon} \widehat{f}_{q, \alpha} 
- \text{Op}_{\varepsilon}( \vert f_{q,\alpha}\vert^{2}) \text{Op}_{\varepsilon}(a \circ \phi)\Vert \leq 
\Vert f_{q,\alpha} \Vert^{2} \big( \Vert a\Vert + \Vert a\circ \phi \Vert\big)\,.$$
Since $\int_{E} \psi(q) \Vert f_{q,\alpha} \Vert^2 \text{d}\nu(q) < \infty$, by properties (P1) and (P2), Sect.~2.1, 
and since the index $\alpha$ has been assumed to take only finitely many values, Lebesgue dominated convergence 
implies that \eqref{limits} holds.

We set
$$\Phi_{\varepsilon, \underline{\psi}_n} := \Phi_{\varepsilon, \psi_0}\circ \cdots \circ \Phi_{\varepsilon, \psi_n}, \quad n=0,1,2, \dots$$
We propose to show by induction that, for an arbitrary $a\in S$ and $n\in \mathbb{N}$,
\begin{equation}\label{norm conv}
\underset{\varepsilon\searrow 0}{\text{lim}}\Vert \Phi_{\varepsilon, \underline{\psi}_{n-1}}(\text{Op}_{\varepsilon}(a)\big)
- \text{Op}_{\varepsilon}\big(\langle \psi_0 \rangle\,\langle \psi_1 \rangle \circ \phi \cdots \langle \psi_{n-1}\rangle \circ
\phi^{n-1} a\circ \phi^{n}\big)\Vert = 0\,.
\end{equation}
In \eqref{limits} this is shown for $n=1$. We now assume that \eqref{norm conv} holds for $n=m-1$. We will use that if 
$\{X_{\varepsilon}\}_{0<\varepsilon \leq \varepsilon_0}$ is a family of operators converging to 0 in norm, as 
$\varepsilon \searrow 0$, then 
\begin{equation}\label{zero}
\underset{\varepsilon\searrow 0}{\text{lim}}\Vert\Phi_{\varepsilon, \underline{\psi}_{m-2}}(X_{\varepsilon})\Vert =0\,.
\end{equation}
Obviously
\begin{align}
\Phi_{\varepsilon, \underline{\psi}_{m-1}}& \big(\text{Op}_{\varepsilon}(a) \big)-
\text{Op}_{\varepsilon}\big(\langle \psi_0\rangle \langle \psi_1\rangle \circ \phi \cdots \langle \psi_{m-1}\rangle \circ \phi^{m-1} a\circ \phi^{m}\big) = \qquad\quad \nonumber \\
= & \Phi_{\varepsilon,\underline{\psi}_{m-2}} \big(\Phi_{\varepsilon, \psi_{m-1}}(\text{Op}_{\varepsilon} (a))- 
\text{Op}_{\varepsilon}(\langle \psi_{m-1}\rangle\, a \circ \phi)\big) + \Phi_{\varepsilon, \underline{\psi}_{m-2}} 
\big(\text{Op}_{\varepsilon}(\langle \psi_{m-1} \rangle\, a \circ \phi)\big) \nonumber \\
- &\text{Op}_{\varepsilon}\big(\langle \psi_0 \rangle \langle \psi_1 \rangle \circ \phi \cdots \langle \psi_{m-2} \rangle 
\circ \phi^{m-2}\big)\cdot \text{Op}_{\varepsilon}\big( \langle \psi_{m-1} \rangle \circ \phi^{m-1} a \circ \phi^{m}\big)
\label{induction step}
\end{align}
We note that the first term on the right side of \eqref{induction step} tends to 0 in norm, by \eqref{limits}, 
and the second term is shown to tend to 0 in norm by using Eq.~\eqref{class lim} and the induction hypothesis. This completes
the induction step proving \eqref{norm conv} for $n=m$.

To complete the proof of Theorem~3.1 we set $a=\langle \psi_n \rangle$ and then use the convergence result 
in \eqref{norm conv} and assumption \eqref{class measure} to show the convergence claimed in \eqref{convergence}.
\hfill$\square$

\medskip
Corollary 3.5 follows from Theorem 3.1 by assuming that the amplitudes $f_{q, \alpha}(\xi)$ only depend on $x$ 
(i.e., are independent of the momentum variable $p$), specializing to translation-invariant instruments, see 
Eqs.~\eqref{transl-invariance} and \eqref{mu measure}, and noticing that the law of $\mathfrak{x}(q_n)$ converges to the law of $x_n + \kappa_n$, 
as $\varepsilon \searrow 0$, where the law of the random variables $\kappa_n$ is given by 
$\vert g(q)\vert^{2} \text{d}q, \,\forall n=0,1,2, \dots$ \hfill$\square$

\section{Proof of Theorem 3.6, and general discussion of results}\label{sec:estim}
In this section we prove Theorem 3.6. As a warm-up, we start with a bare-hands construction of an estimator for $\xi_0$ in the special case
where the particle is freely moving, i.e.,
\begin{equation}\label{free particle}
h_P(x,p)= \frac{p^2}{2m}, \qquad x_n= x_0 + \frac{\tau}{m} n\cdot p_0, \quad \xi_{n} = 
\begin{pmatrix} x_n\\p_0 \end{pmatrix}, \quad n=0,1,2,\dots
\end{equation}
To simplify our notation we choose units such that $(\tau/m)=1$. In Eqs.~\eqref{mu measure} and \eqref{P measure} we have defined the measures 
$$\Lambda(\xi, \text{d}q):= \big| g(q-x) \big|^{2} d^{d}q, 
\qquad P(\xi_0, \text{d}\underline{q}_{\infty}):= \bigtimes_{n=0}^{\infty} \Lambda(\xi_n, \text{d}q_n)\,.$$
In the classical limit, $\varepsilon \searrow 0$, the law of a measurement record 
$\big(Q_n\big)_{n\in \mathbb{N}_0}$ is given by the measure 
\begin{equation}\label{Gibbs measure}
P(\mu_0, \text{d}\underline{q}_{\infty}):= \int_{\Gamma} \text{d}\mu_{0}(\xi)\, P(\xi, \text{d}\underline{q}_{\infty})\,,
\end{equation}
where $\mu_0$ is a probability measure on the space $\Gamma$ of initial conditions.
We temporarily assume that 
$$\text{d}\mu_{0}(\xi)= \delta(x- x_0) \, \delta(p - p_0)\,d^{d}x \, d^{d}p\,, \quad \text{with} \quad \xi=
\begin{pmatrix} x\\p \end{pmatrix}\,.$$
Then the random variables $Q_n$ are independent, because $P(\xi_0, \text{d}\underline{q}_{\infty})$ is a product measure. Moreover, by Corollary 3.5,
$$Q_n=x_0+np_0+\kappa_n\,,$$
where $(\kappa_n)_{n\in \mathbb N_0}$ is a sequence of independent, identically distributed (i.i.d.) random variables 
whose law is given by $|g(q)|^2\text{d}^dq$.
 
The expectation of $Q_n$ is not uniformly bounded in $n$. It diverges as $n\rightarrow \infty$, unless $p_0 =0$. 
It is therefore advantageous to introduce the difference variables 
\begin{equation}\label{diff variiables}
\Delta Q_n := Q_{n+1}-Q_{n}=p_0+(\kappa_{n+1}-\kappa_n),\qquad n\in \mathbb{N}_0\,.
\end{equation}
The random variables $\Delta Q_n$ and $\Delta Q_m$ are independent whenever $|n-m|>1$. It follows that 
$(\Delta Q_{2n+1})_{n\in \mathbb N_0}$ and $(\Delta Q_{2n})_{n\in \mathbb N}$ are two sequences of 
i.i.d. random variables. They have the property that $\mathbb E(\Delta Q_n)=p_0$ and 
$\operatorname{Var}(\Delta Q_n)=2\operatorname{Var}(\kappa_n)<\infty$ for any $n\in \mathbb N$. Hence,
$$\mathbb E(|\Delta Q_n|)<\infty,\quad \mbox{for any }n\in \mathbb N.$$
The strong law of large numbers for i.i.d. random variables applies to $(\Delta Q_{2n+1})_{n\in \mathbb N_0}$ 
and $(\Delta Q_{2n})_{n\in \mathbb N}$ jointly. It follows that,
\begin{equation}\label{eq:free_part_estim_momentum}
\lim_{N\to\infty}\frac{1}{N} \sum_{n=0}^{N-1} \Delta Q_n =\lim_{N\to\infty} \frac{Q_N-Q_0}{N}= p_0, \quad P(\xi_0,\text{d}\underline{q}_\infty)\text{-a.s.}
\end{equation}
Since $\operatornamewithlimits{Var}(\frac{1}{N} \sum_{n=1}^{N} \Delta Q_{2n})=\operatornamewithlimits{Var}(\frac{1}{N} \sum_{n=0}^{N-1} \Delta Q_{2n+1})=\frac{2}{N}\operatornamewithlimits{Var}\kappa_0$, the convergence 
also holds in the norm of $L^2(\mathfrak{Q}, P(\xi_0,\text{d}\underline{q}_\infty))$.
We thus have a consistent estimator 
$$\tilde p(n):=\frac{Q_n-Q_0}{n}$$
of the initial momentum of the particle. We use it to construct an estimator of the initial position of the particle. Assuming for a moment that $\tilde p_n=p_0$ and $Q_k=x_k$ for any $k\in \mathbb N_0$, then $x_0=Q_k-k\tilde p_n$. We thus define,
$$\tilde x_n=\frac1{N_n}\sum_{k=0}^{N_n} \big\{Q_k-k\tilde p_n \big\}$$
with $(N_n)_{n\in \mathbb{N}_0}$ strictly increasing, and $N_n=\mathcal{O}(\sqrt{n})$, as $n$ grows. Then,
$$\tilde x_n-x_0=\frac1{N_n}\sum_{k=0}^{N_n}\big(k(p_0-\tilde p_n) +\kappa_k\big).$$
It follows that
$$\tilde x_n-x_0=\frac{N_n+1}2(p_0-\tilde p_n) + \frac1{N_n}\sum_{k=0}^{N_n}\kappa_k.$$
The second term on the right side vanishes almost surely and in the $L^2$-norm (by the strong law of large numbers 
and because 
$\operatornamewithlimits{Var}(\frac1{N_n}\sum_{k=0}^{N_n}\kappa_k)=\operatornamewithlimits{Var}\kappa_0/N_n$). 
By definition of $\tilde p_n$, the first term is equal to
$$\frac{N_n+1}{2\sqrt{n}}\cdot \frac{np_0-Q_0-Q_n}{\sqrt{n}}=\frac{N_n+1}{2\sqrt{n}} \cdot \frac{\kappa_0-\kappa_n}{\sqrt{n}}.$$
Since $\operatorname{Var}(\frac{\kappa_0-\kappa_n}{\sqrt{n}})=\frac{2\operatorname{Var}(\kappa_0)}{n}$, 
$\lim_{n\to\infty}\frac{\kappa_0-\kappa_n}{\sqrt{n}}=0$ in the $L^2(\mathfrak{Q}, P(\xi_0,\text{d}\underline{q}_\infty))$-norm. 
Moreover, by the strong law of large numbers applied to $(\kappa_n^2)_{n\in \mathbb N_0}$, 
$\frac{\kappa_0-\kappa_n}{\sqrt{n}}$ converges also almost surely to $0$. It then follows from 
the assumed behavior of $N_n$, namely $N_n=O(\sqrt{n})$, that
$$\lim_{n\to\infty} \tilde x_n=x_0,\quad P(\xi_0,\text{d}\underline{q}_\infty)\mbox{-a.s. and in the norm of }
L^2(\mathfrak{Q}, P(\xi_0,\text{d}\underline{q}_\infty))\,.$$
Hence, since the convergence of $(X_n)$ and of $(Y_n)$ implies that the sequence 
$\big((X_n,Y_n)\big)_{n=0,1,\dots}$ converges almost surely and in $L^2$, 
$$\tilde\xi_n:=(\tilde x_n,\tilde p_n)$$
is a consistent estimator of the initial data of the particle, almost surely and in $L^2$, 
hence in probability. More explicitly, in the classical limit $\varepsilon\searrow 0$, $\tilde{\xi}_n$ estimates the 
initial data of the particle more and more precisely, as the number, $n$, of approximate position measurement increases:
$$\lim_{n\to\infty} \tilde \xi_n=\xi_0, \quad P(\xi_0,\text{d}\underline{q}_\infty)\mbox{-a.s.}$$
and in the $L^2(\mathfrak{Q}, P(\xi_0,\text{d}\underline{q}_\infty))$-norm, hence in probability.

Since $\text{d}\mu_0(x, p)=\int_\Gamma \big[\delta(x- x_0) \,\delta(p - p_0) \text{d}\mu_0(x_0, p_0)\big]\text{d}^dx\text{d}^dp$, we can dispose of the assumption that $\text{d}\mu_{0}(x, p)= \delta(x- x_0) \, \delta(p - p_0)\,d^{d}x \, d^{d}p.$
For a non-atomic measure $\text{d}\mu_0$, the initial condition $\xi_0$ becomes random, but Theorem 3.6 continues to hold. 

\medskip
\noindent
{\bf{Remark 5.1.}} \textit{(1) The modest growth of $N_n \sim O(\sqrt{n})$ ensures that the initial momentum estimator
$\tilde p_n$ is close to $p_0$ when used in the estimation of the initial position of the particle. If $N_n$ grew too fast the 
volatility of the initial momentum estimator would prevent the initial position estimator from converging.}\\
\textit{(2) Even if the Hamiltonian of the particle and its initial state $\rho_{0, \varepsilon}$ (as well as the measure 
$\text{d}\mu_0$) are perfectly spherically symmetric an infinitely long sequence of indirect particle-position measurements 
corresponds (almost surely) to a particle motion that breaks the rotational symmetry by singling out an initial value 
of the particle's momentum in a definite (albeit random) direction.}\\
\textit{(3) Arguments similar to the ones described above can be used to
construct an estimator of the initial momentum and position in the direction of the magnetic field 
for the example of a very heavy particle moving in a uniform external magnetic field (see also Sect.~6).}\\

As mentioned in its statement (see Sect.~\ref{survey}), Theorem 3.6 can be extended to quite general quadratic Hamiltonians. 
In the following, we reformulate and prove Theorem 3.6, using a sequence of least-squares estimators $(\tilde \xi_n)$. 
Least-squares estimators minimize the Euclidean distance between the deterministic classical orbit and 
the results of approximate position measurement:
$$\tilde\xi_n:=\operatorname{argmin}_{\xi\in \Gamma} \sum_{k=0}^{2n}\|x_k - Q_k\|^2$$
where $\begin{pmatrix} x_k\\ p_k\end{pmatrix}=\phi^k(\xi)$, and $\|(\cdot)\|$ is the Euclidean norm on 
$\mathbb R^d$. Since $\phi$ is linear, $\tilde\xi_n$ can be found by differentiation.
Let
    $$J=\begin{pmatrix}
        J_{xx} & J_{xp}\\ J_{px} & J_{pp}
    \end{pmatrix}$$
    be the block decomposition of the matrix $J$ corresponding to the symplectomorphism $\phi$ 
    (see Assumption (QF), Sect.~\ref{survey}) with respect to the decomposition 
    $\Gamma=\mathbb R_x^d\oplus \mathbb R_p^d$, 
    and define a matrix $M$ by
	$$M=\begin{pmatrix}
		\mathbf{1}_d &0\\
		J_{xx} & J_{xp}
	\end{pmatrix}.$$
Then
$$\tilde \xi_n:=\left(\sum_{k=0}^n(MJ^{2k})^tMJ^{2k}\right)^{-1}\sum_{k=0}^n (MJ^{2k})^t\begin{pmatrix}
	Q_{2k}\\ Q_{2k+1}
\end{pmatrix}.$$
Of course, this expression holds only if $\sum_{k=0}^n(MJ^{2k})^tMJ^{2k}$ is invertible. 
The hypotheses of the next theorem ensure that this is the case. The theorem asserts 
the consistency of the least-squares estimators (but only in $L^2$; we do not have a 
proof of almost sure convergence).
\begin{theorem}
    Suppose that the hypotheses of Corollary 3.5 and Assumption (QF) of Sect.~\ref{survey} hold. 
    Assume that $\operatorname{spec}J\subset U(1)$ and that $J_{xp}$ is invertible 
    (which implies that $M$ is invertible, too).
    Then the sequence of least squares estimators $(\tilde\xi_n)_{n\in \mathbb N_0}$ converges to $\xi_0$ in the 
    $L^2(\mathfrak{Q}, P(\mu_0,\text{d}\underline{q}_\infty))$-norm. (Thus, convergence also holds in probability.)
\end{theorem}
The assumptions that $\operatorname{spec}J\subset U(1)$ and that $J_{xp}$ is invertible correspond to assumptions {\bf AS} and {\bf AW}, respectively, in paper \cite{BBFF}.\\
Note that, for a free particle with $\tau/m=1$, $\operatorname{spec}J=\{1\}$, $J_{xp}=\mathbf{1}_d$, $M=\begin{pmatrix}
    \mathbf{1}_d & 0\\ \mathbf{1}_d &\mathbf{1}_d
\end{pmatrix}$ and
$$MJ^{2k}=\begin{pmatrix}
    \mathbf{1}_d & 2k\\ \mathbf{1}_d & 2(k+1)
\end{pmatrix}.$$

We end this section with a proof of this theorem and hence of Theorem 3.6. 

\noindent \textit{Proof.} We begin our proof by noting that it suffices to prove the theorem for the measure 
$P(\xi_0,\text{d}\underline{q}_\infty)$, because $P(\mu_0,\text{d}\underline{q}_\infty)$ is a convex combination 
(with respect to the measure $\text{d}\mu_0$) of laws corresponding to deterministic initial data.
	From the hypotheses we infer the equality
	$$\begin{pmatrix}Q_{2n} \\ Q_{2n+1}\end{pmatrix}= MJ^{2n} \xi_0 + \eta_n, \quad \text{for any }\,n\in \mathbb N_0, $$
where $(\eta_n)_{n\in \mathbb N_0}$ is a sequence of $\Gamma$-valued i.i.d.~random variables, with 
$\eta_n=\begin{pmatrix}\kappa_{2n} \\ \kappa_{2n+1}\end{pmatrix}$, and $(\kappa_n)_{n\in \mathbb N_0}$ 
is a sequence of $\mathbb R^d$-valued i.i.d. random variables with law $\vert g(q)\vert^{2}\, d^{d}q$.

	It follows that, for any $n\in \mathbb N_0$,
	$$\tilde \xi_n= \xi_0+\left(\sum_{k=0}^n(MJ^{2k})^{t}MJ^{2k}\right)^{-1}\sum_{k=0}^n (MJ^{2k})^{t}\eta_k\,.$$
	Since the random variables $\eta_n$ are centered, i.i.d. and in $L^2$, we have that $\mathbb E_P(\tilde \xi_n)=\xi_0$ and that there exists a constant $C>0$ such that 
	$$\operatorname{Var}(\tilde \xi_n)\leq C\left(\sum_{k=0}^n(MJ^{2k})^{t}MJ^{2k}\right)^{-1},$$
	for any $n\in \mathbb N_0$.

	Let $\Sigma_n=\sum_{k=0}^n (J^{2k})^{t}J^{2k}$. Since $M$ is real and invertible, $M^{t}M$ is a positive matrix, 
	and there exists a constant $C>0$ such that
	$$\operatorname{Var}(\tilde \xi_n)\leq C \Sigma_n^{-1}.$$
	Hence if we can show that $\Sigma_n \underset{n\to\infty}{\rightarrow} \infty$\footnote{Or, more precisely, that for any $C>0$, there exists a finite $n_0\in \mathbb N_0$ such that for any $n\geq n_0$, $\Sigma_n> C$.} then we conclude that
	the variance of $\tilde{\xi}_n$ converges to 0, as $n\rightarrow \infty$, and therefore $L^2$ convergence holds. 
	The convergence in probability then follows, and the theorem is proven.	
	
	It thus remains to show that $\lim_{n\to\infty}\Sigma_n=\infty$. Since $(\Sigma_n)_{n \in \mathbb{N}_0}$ is an 
	non-decreasing sequence of positive semi-definite matrices of fixed dimension, it suffices to prove that, for an arbitrary 
	$\xi\in \Gamma$ with $\xi \neq 0, \lim_{n\rightarrow \infty} \xi^t\Sigma_n\xi =\infty$. 
	
Let us assume that there exists $\xi\in \Gamma,\, \xi \neq0,$ such that $\xi^t\Sigma_\infty\xi<\infty$. 
This implies that $\lim_{n\to\infty} \|J^{2n}\xi\|=0$. Let $J^2=D+N$ be a decomposition of $J^2$ into a 
diagonalisable matrix $D$ with spectrum in $U(1)$ and a nilpotent matrix $N$, with $[D,N]=0$. (Take 
the Jordan decomposition of $J^2$.) Then
	\begin{equation}\label{eq:conv_Jordan_powers}
		\lim_{n\to \infty}\|(D+N)^n\xi\|=0.
	\end{equation}
	Let $m$ be the smallest integer such that $N^m=0$. Evaluating 
	$(D+N)^n$ explicitly and using that $\sup_n\|D^{-n}\|<\infty$ (since $D$ is diagonalisable and 
	$\operatorname{spec}D\subset U(1)$), one finds that
	Equation~\eqref{eq:conv_Jordan_powers} implies that
	\begin{equation}\label{eq:conv_Jordan_powers_developped}
		\lim_{n\to \infty}\|\sum_{k=0}^{m-1}\binom{n}{k}D^{n-k} N^k\xi\|=0\,,
	\end{equation}
	with the convention that $N^0=\id$. Since, for fixed $k$, $\binom{n}{k}=\frac{n^k}{k!}+\mathcal O(n^{k-1})$, dividing the left hand side by $n^{m-1}$ and taking the limit, we conclude that $N^{m-1}\xi=0$. Repeating this argument for decreasing powers of $n$, one shows by recurrence that $D^n \xi$ tends to 0, as $n\rightarrow \infty$. Since $D$ is invertible and 
	$\sup_n\|D^{-n}\|<\infty$, it follows that $\xi=0$, and we arrive at a contradiction. Thus
	$\Sigma_\infty=\infty,$
	and the theorem is proved.  \hfill{$\square$}

\section{Examples of particle dynamics}
In this last section, we illustrate the general results proven in this paper by discussing standard 
examples of particle dynamics. 
The first two examples have already been discussed in \cite{BBFF}. We allow for 
more general instruments (i.e., more general amplitudes $f_{q, \alpha}$), as compared to \cite{BBFF}. But we study the particle-position measurement process 
only in the vicinity of the classical limit, $\varepsilon \searrow 0$. 
\subsection{Freely moving particle and harmonic oscillators}
We consider $N\geq 1$ particles of mass $m>0$ either freely moving or harmonically coupled. 
The phase-space of this system is given by $\Gamma=\mathbb R_x^{Nd}\oplus \mathbb R_p^{Nd}$. The Hamilton function,
$h_P: \Gamma \to\mathbb R_+$, is given by
\begin{align*}
h_P(x,p) = \frac{1}{2m} (\|p\|^2+x^tOx), \qquad \begin{pmatrix} x\\ p \end{pmatrix}\in \Gamma,
\end{align*}
where $O$ is a real symmetric positive-semi-definite matrix, and $\|(\cdot)\|$ is the euclidean norm on
$\mathbb{R}_{x}^{Nd}$. (For freely moving particles, $O=0$.)

If the particles have a very large mass, as compared to the mass scale of the instrument, it is convenient to replace 
the momentum variables and operators of the particles by their velocities, i.e., $p\to p/m$ (see item 1, 
Eq.~\eqref{CCR'}, Sect.~1.1), and define $\varepsilon:=\hbar/m$. We continue to denote the velocity operator 
by $\hat{p}$. As noted in Eq.~\eqref{CCR'}, we then have that
$$[\hat x_i, \hat p_j]=i\varepsilon \,\delta_{ij}, \qquad i,j=1,\dots, Nd.$$
The unitary time-$\tau$ propagator of the system is given by
$$U_\varepsilon=\exp\Big(-i \frac{\tau}{2\varepsilon} \big(\| \hat p\|_2^2+
\langle \hat x,O \hat x \rangle\big)\Big).$$
Let $J$ be the symplectic $Nd\times Nd$ matrix defined by
$$J=\begin{pmatrix}
	\cos(\sqrt{O}\tau) & \tau\operatorname{sinc}(\sqrt{O}\tau)\\
	-\sqrt{O}\sin(\sqrt{O}\tau) & \cos(\sqrt{O}\tau)
\end{pmatrix}.$$
Then, the classical time-$\tau$ symplectic map on $\Gamma$ determined by the Hamilton function $h_P$ is found to be
$$\phi(\xi)=J\xi\,, \qquad \xi \in \Gamma.$$
Since $h_p$ is quadratic,
$$U_\varepsilon^*\operatorname{Op}_{\varepsilon}(a)U_\varepsilon=\operatorname{Op}_{\varepsilon}(a\circ\phi),$$
for any $a\in S$. Hence Assumption~(SC), Eq.~\eqref{Egorov}, Sect.~2, holds trivially.

\medskip
If $g:\mathbb{R}^{d}\to\mathbb C$ is a Schwartz-space function with the properties that $\int_{\mathbb R^d}|g(q)|^2\text{d} q=1$ 
and $\int_{\mathbb R^d} q|g(q)|^2\text{d} q=0$ then the amplitude $f(\xi;q):\Gamma\times \mathbb R^{Nd}\to \mathbb C$, 
defined by 
$$f(x_1,\dotsc,x_N; p_1,\dotsc, p_N; q_1,\dotsc, q_N)=g(x_1-q_1)\dotsb g(x_N-q_N)\,,$$ 
has properties (P1), (P2) and (P3) stated in Sect.~2.1, with $E=\mathbb R^{Nd}$ and $\text{d}\nu$ 
given by Lebesgue measure on $\mathbb R^{Nd}$. The instrument for approximate position measurements 
corresponding to this choice of an amplitude $f$ is translation-invariant, and Corollary 3.5 holds. 

We now turn to path estimation. We notice that $\operatorname{spec}J\subset U(1)$ and that the block $J_{xp}$ is 
invertible if and only if there does not exist an oscillator eigenfrequency, $\omega$, that after multiplication by $\tau$
is an integer multiple of $\pi$; i.e.,
$$J_{xp}\mbox{ invertible}\iff\nexists\, \omega\in \operatorname{spec}\sqrt{O}\,\,\mbox{ such that }\,\, \omega\tau\in \pi\mathbb 
N.$$
This condition is fulfilled for free particles, since $\operatorname{spec}\sqrt{O}=\{0\}$ and $0\notin \pi\mathbb N$. 
Therefore the hypotheses of Theorem~3.6, concerning the estimation of the initial conditions of the particle trajectories, 
are satisfied. We conclude that, in the large-mass/classical regime of free particles and of harmonic oscillators, and for 
translation-invariant instruments, the initial conditions of the particles can be inferred from a long sequence of approximate position measurements, i.e., from the observed track. 

If the function $g$ is the square root of a \textit{Gaussian density} then we reproduce the setting of paper \cite{BBFF}, 
and the results presented in that paper hold. The assumption that $J_{xp}=\operatorname{sin}(\sqrt{O}\tau)/\sqrt{O}$ 
is invertible corresponds to Assumption~{\bf AW} of \cite{BBFF}.

The next example is inspired by the physics of observing tracks of charged particle in detectors. 
With the purpose of measuring the momentum (or velocity) of charged particles entering a detector, 
one turns on a strong uniform magnetic field pervading the detector.

\subsection{Particle in a strong uniform external magnetic field}
We consider a charged particle in $\mathbb R_x^3$ propagating in a uniform magnetic field $\vec{B}=(0,0,2B)$ 
(perpendicular to the plane $\mathbb R (1,0,0)\oplus \mathbb R (0,1,0)$), with $B>0$. 
We choose units such that the charge of the particle is unity. It follows that the Hamilton function, $h_P$, of
the particle is given by
$$h_P(x,p)=\tfrac1{2m}[(p_1 - B x_2)^2+(p_2 + B x_1)^2]+\tfrac1{2m} p_3^2$$
with $x=(x_1,x_2,x_3)\in \mathbb{R}^{3}_x$, $p=(p_1,p_2,p_3)\in \mathbb{R}^{3}_p$, and
$\Gamma= \mathbb{R}^{3}_{x} \oplus \mathbb{R}^{3}_{p}$; see Eq.~\eqref{phase space}, Sect.~1.2.

We set $\beta:=B/m$, rescale momentum variables as in the previous section ($p\to p/m$) and introduce new variables 
\begin{align}\label{new var}
&y_1=(p_2+\beta x_1)/\sqrt{2\beta}, \quad w_1=(p_1-\beta x_2))/\sqrt{2\beta}, \nonumber\\
& y_2=(p_1+\beta x_2))/\sqrt{2\beta}, \quad w_2=(p_2-\beta x_1))/\sqrt{2\beta}, 
\end{align}
and $ y_3=x_3,\, w_3=p_3$.

One verifies that
$$[\hat w_1,\hat y_1]=\i\varepsilon,\quad [\hat w_2,\hat y_2]=\i \varepsilon\quad \mbox{and}\quad [\hat w_1,\hat w_2]=[\hat y_1,\hat y_2]=0,$$
with $\varepsilon=\hbar/m$. In these new variables
$$h_P(y,w)=m[\tfrac{2\beta} 2(w_1^2+y_1^2)+\tfrac12w_3^2]\,.$$
The time-$\tau$ unitary propagator generated by the quantum Hamiltonian is thus given by
$$U_\epsilon=\exp(-i \tfrac{\tau}{\epsilon}(\tfrac{2\beta} 2(\hat w_1^2+\hat y_1^2)+\tfrac12\hat w_3^2).$$
Since the Hamiltonian is quadratic, 
$$U_\epsilon^*\operatorname{Op}_{\epsilon}(a)U_\epsilon=\operatorname{Op}_\epsilon(a\circ\phi)$$
where $\phi$ is the classical time-$\tau$ symplectic map on $\Gamma$ generated by the Hamilton function $h_P$. 
Since $h_P$ is quadratic, we can determine $\phi$ explicitly: Introducing the symplectic matrices
$$G=\begin{pmatrix}
	\cos(2\beta\tau)&0&0&\sin(2\beta\tau) & 0&0\\
	0&1&0&0&0&0\\
	0&0&1&0&0&\tau\\
	-\sin(2\beta\tau) &0&0& \cos(2\beta\tau)& 0 &0\\
	0&0&0&0&1&0\\
	0&0&0&0&0&1
\end{pmatrix}\quad\mbox{and}\quad P=\frac1{\sqrt{2}}\begin{pmatrix}
\sqrt{\beta}&0&0&0&\frac1{\sqrt{\beta}}&0\\
0&\sqrt{\beta}&0&\frac{1}{\sqrt{\beta}}&0&0\\
0&0&\sqrt{2}&0&0&0\\
0&-\sqrt{\beta}&0&\frac{1}{\sqrt{\beta}}&0&0\\
-\sqrt{\beta}&0&0&0&\frac1{\sqrt{\beta}}&0\\
0&0&0&0&0&\sqrt{2}
\end{pmatrix},$$
we find that $\phi(\xi)=J\xi= P^{-1}GP\xi$.

As in the example of heavy harmonic oscillators, the hypotheses of Theorem 3.6 hold for $\beta$ fixed and 
$\varepsilon\searrow 0$. The limit considered here corresponds to a very heavy particle in a very strong 
magnetic field, with the ratio between particle mass and magnetic field kept constant.

Let $g:\mathbb R^3\to\mathbb C$ be a Schwartz-space function with the properties that $\int_{\mathbb R^3}|g(q)|^2\d q=1$ 
and $\int_{\mathbb R^3}q |g(q)|^2\d q=0$. Choosing amplitudes $f:(\xi,q)\mapsto g(x(\xi)-q)$, one verifies that properties
 (P1), (P2) and (P3) of Sect.~2.1 hold, with $E=\mathbb R^3$ and $\text{d}\nu$ given by the Lebesgue measure on 
 $\mathbb R^3$. The hypotheses of Corollary 3.5 hold.

Concerning the classical path estimation, we note that $\operatorname{spec} J=\operatorname{spec} G=\{1,e^{i2\beta\tau},e^{-i2\beta\tau}\}\subset U(1)$. The upper-right block 
$J_{xp}=\begin{pmatrix}\mathbf{1}_{\mathbb R_x^3}& 0\end{pmatrix}P^{-1}GP\begin{pmatrix}0\\\mathbf{1}_{\mathbb R_p^3}\end{pmatrix}$ 
is given by
$$J_{xp}=\begin{pmatrix}
	\frac{\operatorname{sin}(2\beta \tau)}{2\beta} & -\frac{1-\cos(2\beta\tau)}{2\beta} & 0\\
	\frac{1-\cos(2\beta\tau)}{2\beta}& \frac{\operatorname{sin}(2\beta \tau)}{2\beta} &0\\
	0&0&\tau
\end{pmatrix}.$$
If $\beta\tau\notin \pi\mathbb N$ then $J_{xp}$ is invertible, and the assumptions of Theorem 3.6 hold. Thus, 
in the limit of a large particle mass and a large magnetic field, the initial momentum and position of the charged 
particle can be inferred from the particle track in the detector. 

Taking $g$ to be the square root of a Gaussian density we recover the setting of paper \cite{BBFF}, 
and the results presented there apply.

\medskip
The dynamics of the next example is not described by a linear symplectic matrix on phase space and hence does not fit 
into the setting of \cite{BBFF}.

\subsection{Particle in a smooth external potential}
Let $S=\{ a\in C^{\infty}(\Gamma): \|\partial^{\alpha}a\|_\infty<\infty, \forall \alpha\}$. We say 
$V:\mathbb R_+^*\times \Gamma\to \mathbb R$ is a semi-classical potential if for some $m_0>0$, 
there exists a sequence, $(V_j)_{j\in \mathbb N_0}$, of functions in $S$ such that, for any $N\in \mathbb N_0$ and 
an arbitrary multi-index $\alpha$, there exists a constant $C>0$ such that, for all $m>m_0$,
$$\sup_{\xi\in \Gamma}\left|\partial_\xi^\alpha\left(V(m,\xi)-\sum_{j=0}^Nm^{-j+1}V_j(\xi)\right)\right|\leq Cm^{-N}.$$
Assume $V$ is a semi-classical potential. Then $\lim_{m\to\infty}\|\frac1m V(m,\cdot)- V_0\|_\infty=0$. In particular 
if $V$ is independent of $m$, $V_0=0$.

According to Egorov's theorem (see, e.g., \cite[Theorem 1.2]{BR}), setting $\varepsilon=\hbar/m$ and performing 
the same rescaling of the momentum variables as in the previous two examples ($p\to p/m$), 
the time-$\tau$ propagator given by
$$U_\epsilon=\exp\left(-i \frac{\tau}{\epsilon}\operatorname{Op}_\epsilon\left(\| p \|_2^2+\tfrac1{m}V(m,\cdot)\right)\right)$$
is a well defined unitary operator, and Assumption~(SC), Eq.~\eqref{Egorov}, of Sect.~2 holds, with $\phi$ 
determined by the classical Hamilton function
$$h_{P}(x,p) = \tfrac12\|p\|^2+V_0(x)\,.$$
Ultraviolet regularized versions of the gravitational potential or the Lennard-Jones potential (with strength proportional 
to the mass of the particle) are examples of semi-classical potentials. However, the definition given above characterizes 
a considerably more general class of potentials that scale like $V(m,x)\sim mW(x)$ as the mass $m$ becomes large, for 
a smooth effective potential $W$. 

Since the dynamics of this example is non-linear, our results on the estimation of initial conditions of particle trajectories 
do not apply directly. However, using results on classical and quantum-mechanical scattering theory in 
external potentials with rapid fall-off at $\infty$, one expects to be able to extend these results to examples 
of the kind considered here. Further study of details is desirable.

\appendix

\section{Weyl quantization and semi-classical analysis}\label{app:semi-classical}
In this appendix, we review Weyl quantization for appropriate spaces, $S$, of functions on phase space $\Gamma$ 
and study the validity of Assumption (SC) of Sect.~2 concerning the classical limit.

For linear functionals, $l:\xi\mapsto \xi_l^{t}\cdot \Omega \xi$, we define $\Op_\epsilon(l)=\xi_l^{t}\cdot \Omega \hat \xi$, and, for quadratic functions 
$c:\xi\mapsto \xi^{t}\cdot C\xi$, with $C$ symmetric, we define $\Op_\epsilon(c)=\hat{\xi}^{t}\cdot C \hat\xi$.
In the next subsection we define Weyl quantization for some spaces of bounded functions.

\subsection{Function spaces}
\begin{definition}[$S_k$]
Let $\mathcal M(\Gamma)$ be the vector space of finite, complex Borel measures on phase space $\Gamma$. We define
a function space $S_0$ as the image of $\mathcal M(\Gamma)$ under inverse Fourier transformation, 
$\mathcal F^{-1}$:
\begin{equation}\label{Borel meas}
a\in S_0\implies \exists \mu_a\in \mathcal M(\Gamma), \,\ \mbox{ such that }\,\ a(\xi)=\int_\Gamma \e^{\i \zeta^{t}\cdot 
\Omega \xi}\d\mu_a(\zeta)\,,\,\ \mbox{ with }\,\ |\mu_a|(\Gamma)<\infty.
\end{equation}
For $k\in \nn$, we define the space $S_k$ to be the subspace of $S_0$ with the property that $a\in S_k$ implies 
$\int_\Gamma \|\zeta\|^k\, \d|\mu_a|(\zeta) <\infty$.
We equip $S_k$ with the norm $\|a\|_{TV(k)}=\int_\Gamma(1+\|\zeta\|)^k \d|\mu_a|(\zeta)$ (using the 
convention that $x^0=1$). This turns $S_k$ into a Banach space. When equipped with point-wise multiplication 
and point-wise complex conjugation, $a\mapsto\bar a$, $S_k$ becomes a commutative, normed $^{*}$-algebra.

We define $S_\infty=\bigcap_{k\in \nn} S_k$ and use the shorthand $\|a\|_{TV}=\|a\|_{TV(0)}$.
\end{definition}
Note that $S_{k+1}\subset S_k$, for all $k=0,1,\dotsc,\infty$. Let $\mathcal{S}(\Gamma)$ be the Schwartz space of 
test functions on phase space $\Gamma$ (see Sect.~2). Since $\mathcal F\mathcal S(\Gamma)\subset\mathcal S(\Gamma)$, 
$\mathcal S(\Gamma)\subset S_\infty$. But $S_\infty$ is significantly larger than $\mathcal S(\Gamma)$. 
For example, the constant functions belong to $S_\infty$, and if $f\in \mathcal S(\rr_x^d)$ then $a:(x,p)\mapsto f(x)$ 
is an element of $S_\infty$. 

For any $k\in \nn$, $a\in S_k$ entails that $a$ is $k$ times continuously differentiable, with bounded derivatives. Hence $S_\infty\subset \{a\in C^{\infty}(\Gamma) : \|\partial^\alpha a\|_\infty<\infty, \forall \mbox{ multi-index } \alpha\}$. The converse inclusion does not hold a priory.

For any $a\in S_0$, we define its \textit{Weyl quantization} using a bounded bilinear form on 
$\mathcal H_P \times \mathcal H_P$ and appealing to the Riesz's representation theorem, as 
described in Sect.~2.
\begin{proposition}\label{prop:def_quantization}
Suppose that $a\in S_0$. Then 
\begin{align*}
	B_a:\mathcal H_P\otimes \mathcal H_P:&\to \cc\\
	(\Phi,\Psi)&\mapsto \int_\Gamma\langle\Phi,W(\zeta)\Psi\rangle \d \mu_a(\zeta)\,,
\end{align*}
is a well defined bounded sesquilinear form. (Here $W(\zeta) :=\text{exp}\big[i(\zeta^{t}\Omega \hat{\xi}\,)\big]$ is
the Weyl operator associated with $\zeta \in \Gamma$, see \eqref{W Op}, Sect.~\ref{quantization}; in Eq.~\eqref{sesqui}, Sect.~\ref{quantization}, $B_a$ has been denoted by $B_{\varepsilon}(a |\cdot, \cdot)$; the measure $ \text{d}\mu_a$ 
is as in \eqref{Borel meas}.)

Moreover, there exists a unique operator $\Op_\varepsilon(a)\in \mathcal B(\mathcal H_P)$ such that
\begin{equation}\label{operator}
B_a(\Phi,\Psi)=\langle\Phi,\Op_{\varepsilon}(a)\Psi\rangle\quad \mbox{and}\quad \|\Op_\varepsilon(a)\|\leq \|a\|_{TV}.
\end{equation}
\end{proposition}
\textit{Proof.} The Riesz representation theorem tells us that it suffices to prove that $B_a$ is a well defined bounded sesquilinear 
form on $\mathcal H_P\times \mathcal H_P$ with a norm smaller than $|\mu_a|(\Gamma)$, in order to conclude that 
a bounded operator $Op_{\varepsilon}(a)$ satisfying \eqref{operator} exists. We have noted in Sect.~\ref{quantization}, 
above Eq.~\eqref{sesqui}, that  the function $\zeta \mapsto \langle\Phi,W(\zeta)\Psi\rangle$ is continuous in 
$\zeta$, and, since Weyl operators are unitary, $|\langle\Phi,W(\zeta)\Psi\rangle|\leq \|\Phi\|\|\Psi\|$. 
Thus $\langle\Phi,W(\cdot)\Psi\rangle$ is continuous and bounded,
hence $\mu_a$-integrable. It follows that $B_a$ is well defined. It is sesqulinear, because the integral with respect 
to $\mu_a$ is linear, and $(\Phi,\Psi)\mapsto \langle\Phi,W(\zeta)\Psi\rangle$ is sesquilinear. 
Finally, from the definition of $B_a$,
$$|B_a(\Phi,\Psi)|\leq\int_\Gamma|\langle\Phi,W(\zeta)\Psi\rangle|\d|\mu_a|(\zeta)\leq |\mu_a|(\Gamma)\  \|\langle\Phi,W(\cdot)\Psi\rangle\|_\infty\leq |\mu_a|(\Gamma)\ \|\Phi\| \cdot \|\Psi\|.$$
It follows that $B_a$ is bounded by $|\mu_a|(\Gamma)$ and the proposition is proved.  \hfill{$\square$}

This definition of quantization can be extended to unbounded functions, $a$, in which case the sesquilinear form 
$B_a$ is defined only on a dense subspace of $\mathcal H_P \times \mathcal{H}_P$; see \cite[\S VIII.6]{reed1980methods}. 
In particular, for an arbitrary linear function $l:\Gamma\to \rr$ and a function $a\in S_k$, $\Op_\varepsilon(la)$ is well defined.

Next, we prove that $\mathcal A_k=\{\Op_\epsilon(a): a\in S_k\}$ is a $^{*}$-algebra.
\begin{proposition}\label{prop:algebra}
For all functions $a,b\in S_k$, where $k \in \mathbb{N}_0$ is arbitrary, $\Op_{\varepsilon}(a)^*=\Op_{\varepsilon}(\bar a)$, 
$\Op_{\varepsilon}(z\cdot a)= z \cdot Op_{\varepsilon}(a), \forall z\in \mathbb{C}$, $\Op_\varepsilon(a)+\Op_\varepsilon(b)=\Op_\varepsilon(a + b)$, and 
$$\Op_\varepsilon(a)\cdot \Op_\varepsilon(b)=\Op_\varepsilon(a\star b),$$
where the star product, $a\star b$, of $a$ and $b$ is defined by
$$ \int_\Gamma f(\zeta)\d\mu_{a\star b}(\zeta)=\int_{\Gamma^2} f(\zeta_1+\zeta_2) \e^{-\i\frac\varepsilon2 \zeta_1^t\Omega\zeta_2}\d\mu_a(\zeta_1)\d\mu_b(\zeta_2)$$
for an arbitrary bounded continuous function $f$. \\
One has that $a\star b\in S_k$, and $(\mathcal A_k, +, \cdot, *)$ is a $^{*}$-algebra of bounded operators.
\end{proposition}
\noindent \textit{Proof.} Clearly, $a\mapsto B_a$ is linear, so that it follows from the uniqueness of the operator representative, 
$Op_{\varepsilon}(a)$, that $\Op_{\varepsilon}(z\cdot a)= z \cdot Op_{\varepsilon}(a), z \in \mathbb{C},$ and 
$\Op_\varepsilon(a)+\Op_\varepsilon(b)=\Op_\varepsilon(a+b)$. 
Furthermore, since $W(\zeta)^*=W(-\zeta)$, we have that $\overline{B_a(\Phi,\Psi)}=B_{\bar a}(\Psi,\Phi)$. 
Uniqueness of the operator representative of $B_a$ then implies that $\Op_\varepsilon(a)^*=\Op_{\varepsilon}(\bar a)$.
Finally, we show that $\Op_\varepsilon(a)\cdot \Op_{\varepsilon}(b)=\Op_\varepsilon(a\star b)$, with $a\star b\in S_k$. 
The definition of $\star$, combined with the obvious inequality $1+x+y\leq (1+x)(1+y)$, for $x, y$ non-negative
and the triangle inequality, implies that
$$\int_\Gamma  (1+\|\zeta\|)^k\d|\mu_{a\star b}|(\zeta)\leq\int_{\Gamma^2}(1+\|\zeta_1\|+\|\zeta_2\|)^k\d|\mu_a(\zeta_1)|\d|\mu_b(\zeta_2)|\leq\|a\|_{TV(k)}\|b\|_{TV(k)}.$$
Thus $\mu_{a\star b}$ is a finite measure, and, for $a\star b\in S_k$, $\|a\star b\|_{TV(k)}\leq \|a\|_{TV(k)}\|b\|_{TV(k)}$.
Since $\mathcal H_P$ is separable, it has a countable orthonormal basis $\{\chi_n\}_{n\in \nn}$, with 
$\sum_{n\in \nn} |\chi_n\rangle\langle\chi_n|=\id_{\mathcal H_P}$. By definition,
$$\langle \Phi,\Op_{\varepsilon}(a)\cdot \Op_\varepsilon(b)\Psi\rangle=\sum_{n\in \nn} B_a(\Phi,\chi_n)\cdot B_b(\chi_n,\Psi).$$
Fubini's theorem then implies that
$$\langle \Phi,\Op_{\varepsilon}(a)\cdot \Op_\varepsilon(b)\Psi\rangle=\int_{\Gamma^2}\langle \Phi,W(\zeta_1)\cdot W(\zeta_2)\Psi\rangle\, \d\mu_a(\zeta_1)\,\d\mu_b(\zeta_2).$$
The Weyl relations (see Eq.~\eqref{WR}, Sect.~2) then yield
$$\langle \Phi,\Op_{\varepsilon}(a)\cdot \Op_\varepsilon(b)\Psi\rangle=\int_{\Gamma^2}\langle \Phi,W(\zeta_1+\zeta_2)\Psi\rangle\, \e^{-\i\frac\epsilon2\zeta_1^t\Omega\zeta_2}\,\d\mu_a(\zeta_1)\,\d\mu_b(\zeta_2),$$
and the proposition is proved. \hfill{$\square$}

\vspace{0.15cm} When equipped with the $\star$ product, instead of the point-wise product, the space $S_k$ is a $C^*$-algebra. 
But the algebra $\mathcal A_k$ is \textit{not} closed in the \textit{operator norm} on $\mathcal H_P$. We denote the 
norm closure of $\mathcal A_k$ by $\mathcal C$, i.e., $\mathcal{C}$ is the smallest $C^*$-algebra such that 
$\mathcal A_k\subset \mathcal C$. Since $\|\cdot\|_{TV}$ dominates the operator norm, and since $S_k$ is 
norm-dense in $S_0$, $\mathcal C$ is independent of $k$. The weak closure $\mathcal C''$ of 
$\mathcal C$ is actually Weyl's CCR algebra. 

\subsection{Classical limit}
Using our definition of quantization, we can apply the Lebesgue dominated convergence theorem to prove existence of
the classical limit, as stated in Eq.~\eqref{class lim} and Assumption (SC) of Sect.~2.

\begin{proposition}\label{prop:classical_limit_product}
Let $a,b\in S_0$. Then,
$$\lim_{\varepsilon\downarrow 0} \|\Op_\varepsilon(a)\cdot \Op_\varepsilon(b)-\Op_\varepsilon(a\cdot b)\|=0.$$
\end{proposition}
\noindent \textit{Proof.} By Proposition~\ref{prop:algebra},
$$\Op_\varepsilon(a)\cdot \Op_\varepsilon(b)=\Op_\varepsilon(a\star b).$$
We note that $\mu_a\ast\mu_b$ is the finite measure whose Fourier transform is given by pointwise multiplication of 
$a$ with $b$, i.e., $\mu_a*\mu_b=\mu_{a\cdot b}$, and that
$$|\mu_{a\star b}-\mu_{a\cdot b}|(\Gamma)\leq\int_{\Gamma^2} \left|\e^{-\i\frac{\epsilon}2\zeta_1^t\Omega\zeta_2}-1
\right|\d|\mu_a|(\zeta_1)\d|\mu_b|(\zeta_2)\leq 2 \vert \mu_a\vert (\Gamma) \vert \mu_{b}\vert(\Gamma).$$
The proof of the proposition is then completed by invoking Lebegue's dominated convergence theorem 
and the bound $\Op_\varepsilon(c)\leq |\mu_c|(\Gamma)$, \,$\forall\,c\in S_0$.  \hfill{$\square$}

\vspace{0.15cm}Since $S_k\subset S_0$, this proposition shows that, for any $k=0,1,\dotsc,\infty$, $S_k$ is an appropriate choice 
of a function space $S$ in our quantization procedure.

\medskip
Next, we prove that, for a Hamilton function $h_P=h_0+V$, where $h_0$ is a real polynomial in $\xi\in \Gamma$ 
of degree at most $2$ and $V$ is a potential belonging to $S_1$, Assumption (SC), Sect.~\ref{quantization}, in particular 
Eq.~\eqref{Egorov} hold.
\begin{proposition}
	Let $t\mapsto \phi^t, t\in \mathbb{R},$ be the symplectic flow generated by a Hamilton function 
	$h_{P}(\xi) = h_0(\xi)+V(\xi), \xi \in \Gamma$,
	 where $h_0(\xi)$ is a real polynomial in $\xi$ of degree at most $2$, and $V\in S_1$ is real. We assume that for an arbitrary
	 $a\in S_1$, the function $t\mapsto\|a\circ\phi^t\|_{TV(1)}$ is uniformly bounded on compact subsets of $\mathbb{R}$.
	 We set $U_\varepsilon:=\exp\left(-i\frac{\tau}{\varepsilon} \Op_\varepsilon(h_P)\right)$.
	 
	 Then, choosing $S=S_1$ and setting $\phi:=\phi^{\tau}$, Assumption (SC) of Sect.~2 holds. \end{proposition}

\noindent \textit{Proof.} By assumption, we have that $\phi(S_1)\subset S_1$. By definition,  
	$$\partial_t \big(a\circ\phi^{\tau-t}\big)=-\{a,h_P\}\circ\phi^{\tau-t}=-\{a\circ\phi^{\tau-t},h_P\},$$ 
	for an arbitrary $a\in S_1$, where $(a,b)\mapsto\{a,b\}$ is the Poisson bracket. 
	
	For $a,b\in S_1$, we define $\{\{a,b\}\}:=-\frac i\epsilon (a\star b-b\star a)$. Using Duhamel's trick and the 
	anti-symmetry of the Poisson bracket, it follows from Proposition~\ref{prop:algebra} that
	$$U_{\varepsilon}^*\Op_\varepsilon(a)U_\varepsilon - \Op_\varepsilon(a\circ\phi)=
	\int_0^{\tau} \e^{\frac{i}{\varepsilon}t\Op_\varepsilon(h_P)}\Op_\varepsilon(\{\{a\circ\phi^{\tau-t},V\}\} 
	-\{a\circ\phi^{\tau-t},V\})\e^{-\frac i\varepsilon t\Op_\varepsilon(h_P)} \d t.$$
	Here we have used that, since $h_0$ is at most of degree $2$, one has that $-\frac{i}{\varepsilon}[\Op_{\varepsilon}(b),
	\Op_\varepsilon(h_0)]=\Op_\varepsilon(\{b,h_0\})$, for any $b\in S_1$ (see \cite[Theorem~10.13]{DerGerard} for example).
	The above identity implies that
	\begin{align}\label{eq:duhamel_bound}
		\|U_{\varepsilon}^*\Op_\varepsilon(a)U_\varepsilon - \Op_\varepsilon(a\circ\phi)\|\leq 
		\int_0^{\tau}\|\{\{a\circ\phi^{\tau-t},V\}\} -\{a\circ\phi^{\tau-t},V\}\|_{TV}\,\d t.
	\end{align}
	For arbitrary $a,b\in S_1$, $g_{a,b}=\{\{a,b\}\} -\{a,b\}$ is the inverse Fourier transform of
	$$\tilde g_{a,b}:\zeta\mapsto\int_{\Gamma^2}\delta(\zeta-(\zeta_1+\zeta_2))\Big[\frac{\operatorname{sin}(\tfrac\epsilon 2 \zeta_2^t\Omega \zeta_1)}{\tfrac\epsilon 2 \zeta_2^t\Omega \zeta_1}-1\Big]\,(\zeta_2^t\Omega\zeta_1)\,\d \mu_a(\zeta_1)
	\,\d \mu_b(\zeta_2).$$
	It follows that $\|g_{a,b}\|_{TV}\leq 2\|a\|_{TV(1)}\|b\|_{TV(1)}$. Hence, by hypothesis, the integrand of the integral on
	the right side of \eqref{eq:duhamel_bound} is uniformly bounded in $t$, hence integrable on the interval $[0,\tau]$. 
	Since $\lim_{x\to 0}\frac{\operatorname{sin}(x)}{x}=1$, Lebesgue's dominated convergence theorem implies
	that $\tilde{g}_{a,b}$ converges to $0$, as $\varepsilon$ tends to 0, and hence 
	$$\underset{\varepsilon \searrow 0}{\text{lim}}\, \|U_{\varepsilon}^*\Op_\varepsilon(a)U_\varepsilon - \Op_\varepsilon(a\circ\phi)\| 
	= 0.$$
	This completes the proof of the proposition. \hfill{$\square$}
	
\vspace{0.15cm} We remark that the assumption that $t\mapsto \|a\circ\phi^t\|_{TV(1)}$ is uniformly bounded on compact 
sets of $\mathbb{R}$ holds if $V=0$.

\begin{center}
-----
\end{center}

\bigskip

\noindent
Tristan Benoist, Institut de Math\'ematiques de Toulouse, UMR5219, Universit\'e de Toulouse, CNRS, UPS, F-31062 Toulouse Cedex 9, France;\\ \href{mailto:tristan.benoist@math.univ-toulouse.fr}{tristan.benoist@math.univ-toulouse.fr}.
\\[0.3em]
Martin Fraas, Department of Mathematics, University of California, Davis, Davis, CA, 95616, USA;\\ \href{mailto:martin.fraas@gmail.com}{martin.fraas@gmail.com}.
\\[0.3em]
J\"urg Fr\"ohlich, ETH Zurich, Department of Physics, HIT K42.3, CH-8093 Zurich, Switzerland;\\\href{mailto:juerg@phys.ethz.ch}{juerg@phys.ethz.ch}

\end{document}